\documentclass[showpacs, pra,onecolumn,preprintnumbers ,amsmath, amssymb, superscriptaddress, aps]{revtex4-2}
\usepackage{color}
\usepackage{amsmath,amssymb}
\usepackage{pifont}% http://ctan.org/pkg/pifont
\usepackage{amssymb}  % More symbols
\usepackage{bbold}
\usepackage{float}
\usepackage{subfloat}

\usepackage[caption=false]{subfig}
\usepackage{tikz}
\usepackage{makecell}
\usepackage{subfig}
\usepackage{pifont}   % Ding symbols
\usepackage{graphicx} % Include figure files
\graphicspath{{Figure2s/}}
\usepackage{dcolumn}  % Align table columns on decimal point
\usepackage{bm}       % bold math
\usepackage{multirow} % Table functions
\usepackage{placeins}% For making floats not move around everywhere
\usepackage[colorlinks]{hyperref}
\usepackage{mathtools}
\usepackage{appendix}

\newcommand{\rr} {\color{red}}

\newcommand{\bb} {\color{blue}}

\begin{document}
	\title{Transmission in strained graphene subjected to laser and magnetic fields}
	\date{\today}
	\author{Hasna Chnafa}
	\affiliation{Laboratory of Theoretical Physics, Faculty of Sciences, Choua\"ib Doukkali University, PO Box 20, 24000 El Jadida, Morocco}
	%   \email{el-hassouny.j@ucd.ac.ma}
	\author{Miloud Mekkaoui}
	\affiliation{Laboratory of Theoretical Physics, Faculty of Sciences, Choua\"ib Doukkali University, PO Box 20, 24000 El Jadida, Morocco}
	\author{Ahmed Jellal}
	\email{a.jellal@ucd.ac.ma}
	\affiliation{Laboratory of Theoretical Physics, Faculty of Sciences, Choua\"ib Doukkali University, PO Box 20, 24000 El Jadida, Morocco}
	\affiliation{Canadian Quantum  Research Center,
		204-3002 32 Ave Vernon,  BC V1T 2L7,  Canada}
\author{Abdelhadi Bahaoui}
\affiliation{Laboratory of Theoretical Physics, Faculty of Sciences, Choua\"ib Doukkali University, PO Box 20, 24000 El Jadida, Morocco}	
%	\author{El Houssine Atmani}
%	%   \email{}
%	\affiliation{Laboratory of Nanostructures and Advanced Materials, Mechanics and Thermofluids, Faculty of Sciences and Techniques, Hassan II University, Mohammedia, Morocco}
	
	
	\begin{abstract} 
We investigate the effect of strain along armchair and zigzag directions on electrical transport in graphene through a magnetic barrier and a linearly polarized electromagnetic wave. In the context of Floquet theory, the eigenvalues and related eigenspinors are calculated analytically. The transmission probabilities are expressed as a function of different parameters using the transfer matrix approach and boundary conditions at two interfaces with current densities. We see that as the barrier width and incident energy change, the transmission via the center band oscillates less at zero strain. The transmission across the first sidebands begins at 0 and follows the pattern of a sinusoidal function that grows with increasing barrier width and becomes nearly linear for larger incident energy. When the strain magnitude is activated, the number of oscillations in all transmission channels drops marginally in the armchair direction but increases dramatically in the zigzag direction.  
The behavior of the total transmission is found to be comparable to that of the central band, with the exception that it exhibits a translation to the up. The suppression of Klein tunneling at normal incidence is another result seen in all strain settings. 	 

	\end{abstract}

	\maketitle
	
%%%%%%%%%%%%%%%%%%%%%%%%%%%%%%%%%%%%%%%%%%%%%%%%%%
    \section{Introduction}

Because of its extraordinary physical properties, graphene has quickly become a fascinating material that has attracted scientists since its discovery in 2004, \cite{s1,s2}. It does, in fact, have a higher electrical mobility than $2.105$ \text{cm$^{2}$}\text{V$^{-1}$}\text{s$^{-1}$} \cite{s3,s4}, 
 a good flexibility \cite{sb}, a Hall effect \cite{s5,s6,s7}, an elastic strain engineering \cite{s8,s9,s10,s11,saz,saz1,saa,saag,ssa2,ssa3,ssa4,adz}, a Klein tunneling \cite{s11a,s11b}, and other properties.  
The particularity of graphene does not halt at these properties. It is shown that the electrons move in graphene $300$ times slower than light.  Also, graphene is almost transparent, absorbing $2.3\%$ of white light \cite{s14a} 
and has an effective Young's modulus of $\sim1$ \text{TPa} \cite{s25, ah1}.
The conditions to which the electrons are subjected are described mathematically by the Dirac-Weyl equation, as for a massless relativistic particles \cite{s13,s14,s143}.

When a mechanical constraint, such as a tension \cite{s144} or a compression \cite{s145}, is applied to graphene, it undergoes uniform deformation, which alters its properties and improves its technological applications. Another way has been suggested to realize this kind of strain: depositing graphene on a transparent, flexible substrate made of polyethylene terephthalate (PET), and then stretching the PET in one direction \cite{s133}.  This unusual range of elastic response opens a new opportunity to explore the changes induced by the mechanical constraints on the electronic properties of graphene. 
It is found that on account of distortion, the structure of graphene becomes asymmetric and causes asymmetric interactions between the electrons of its three nearest neighbors and the electrons in the sublattices. Therefore, different changes are made to the three positions of the electrons' nearest neighbors and the three hoping energies. As a result, the Dirac points change and a spectral gap opens between the conduction and valence bands \cite{s8,s10}, resulting in Dirac fermions with unequal Fermi velocities $v_x\neq v_y$ \cite{sg1,Sz}. Lately, it has been shown that the strain in graphene is also achieved non-uniformly as a pseudomagnetic field \cite{ah441,ah5}.
Experimentally, it has been discovered that the local field of a non-uniform deformation can create a pseudo-magnetic field larger than $300$ \text{T} \cite{ah3,ah4,ah44}.

Furthermore, researchers are currently interested in electron processes associated with dressing fields, which are being investigated in a number of systems 
{\bb\cite{sx,sx1,sx4,sx5,sx6,sx7,sx8,sx9,Nag2019,sx10,sx11,XX3}}.
Moreover, the production of dynamical gaps in the spectra of Dirac electrons {\bb\cite{XX,XXZ,ggt}}, and the removal of the Klein tunneling effect by strong radiation {\bb\cite{XX1,qqa,str1,dr1}} are among the first noteworthy results obtained with graphene dressed by the monochromatic field.
Many theoretical investigations of electronic transport 
have demonstrated and shown that the behavior of the transmission probabilities is highly influenced by the amplitude and frequency of laser light {\bb\cite{qqa,str1,dr1,A1, Xu51,Xu5,strain1,Ipsita2021,dr1}}. Although various works carried out on laser-aided graphene have been published, the existence of strain impact under a laser field through a delta-function magnetic barrier is an attractive problem to be investigated, which constitutes the focus of our paper.

In this work, we aim to study the influence of strain along armchair and zigzag directions on the tunneling spectra in graphene laser-magnetic barriers. Our system consists of three regions, with the intermediate one being subjected to uniaxial strain and a magnetic field and irradiated by a linearly polarized electromagnetic wave.
The energy spectrum and their solutions for each region are obtained by solving the Dirac equation. Afterwards, the conditions for the limits and the transfer matrix approach are used to calculate the transmission probability.  The effect of strain on transmission behaviors in the armchair and zigzag directions will be numerically studied. As a result, we show that transmission probability oscillations via graphene magnetic barriers vanish slightly for armchair deformation but reappear dramatically for zigzag deformation.
We conclude that the strain magnitude can be used to influence the tunneling features of our system.

The manuscript of the proposed paper is organized as follows. In  Sec. \ref{TFor}, we mathematically formulate our problem and identify the eigenspinors of the energy spectrum. Applying the continuity condition at two interfaces with a transfer matrix in order to analyze the transmission probability for all channels in Sec. \ref{Tran}. Numerically, we examine and explain our results by displaying several plots under various condition in Sec. \ref{ANA}. 
We close by concluding our results.

%%%%%%%%%%%%%%%%%%%%%%%%%%%%%%%%%%%%%%%%%%%%%%%%%%
\section{Theoretical formulation}\label{TFor}

\begin{figure}[H]
        \centering
            \subfloat[]{\includegraphics[width=0.4\linewidth, height=0.21\textheight]{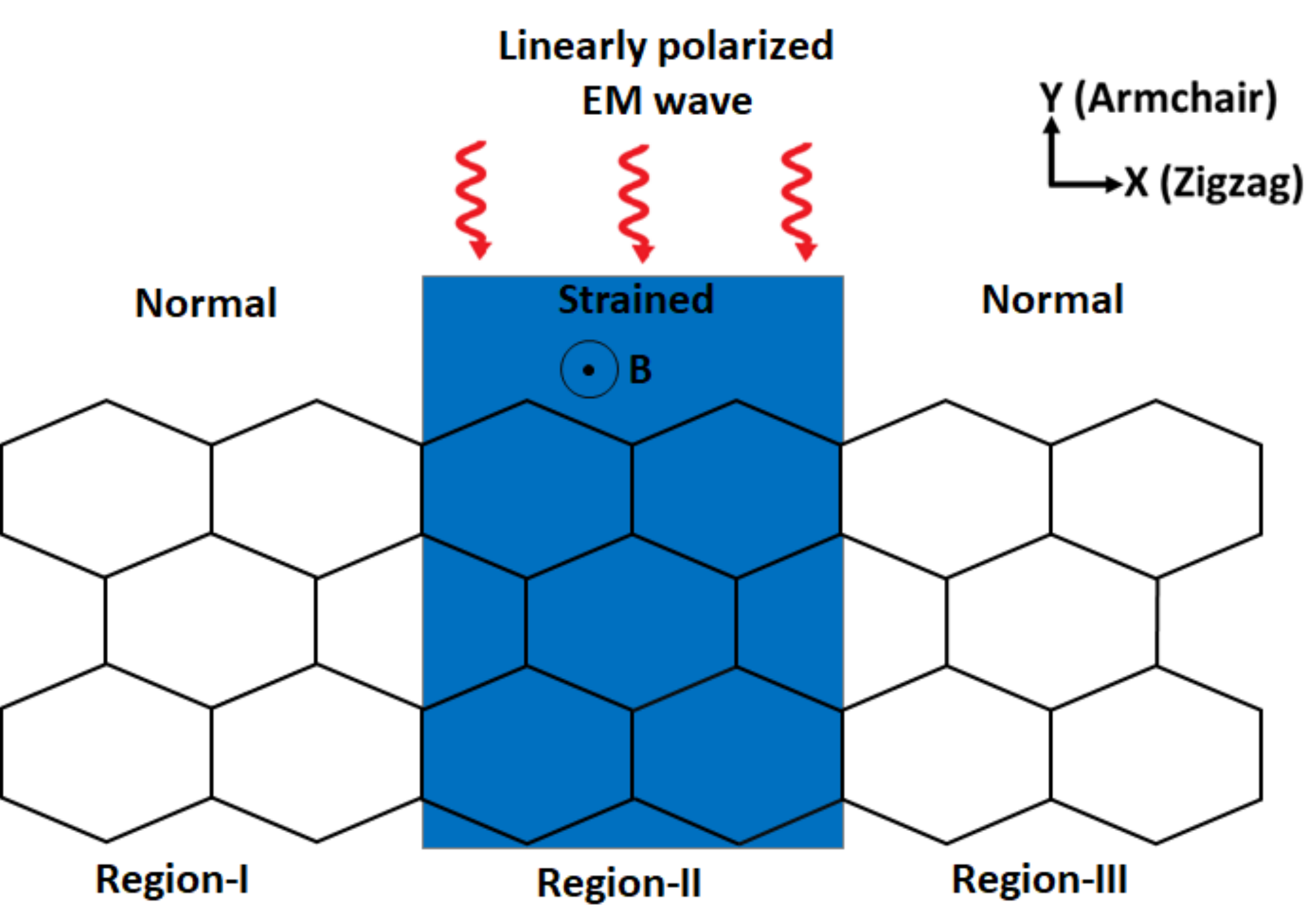}}\ \ \ \
            %\hspace{0.5cm}
        \subfloat[]{\includegraphics[width=0.4\linewidth, height=0.21\textheight]{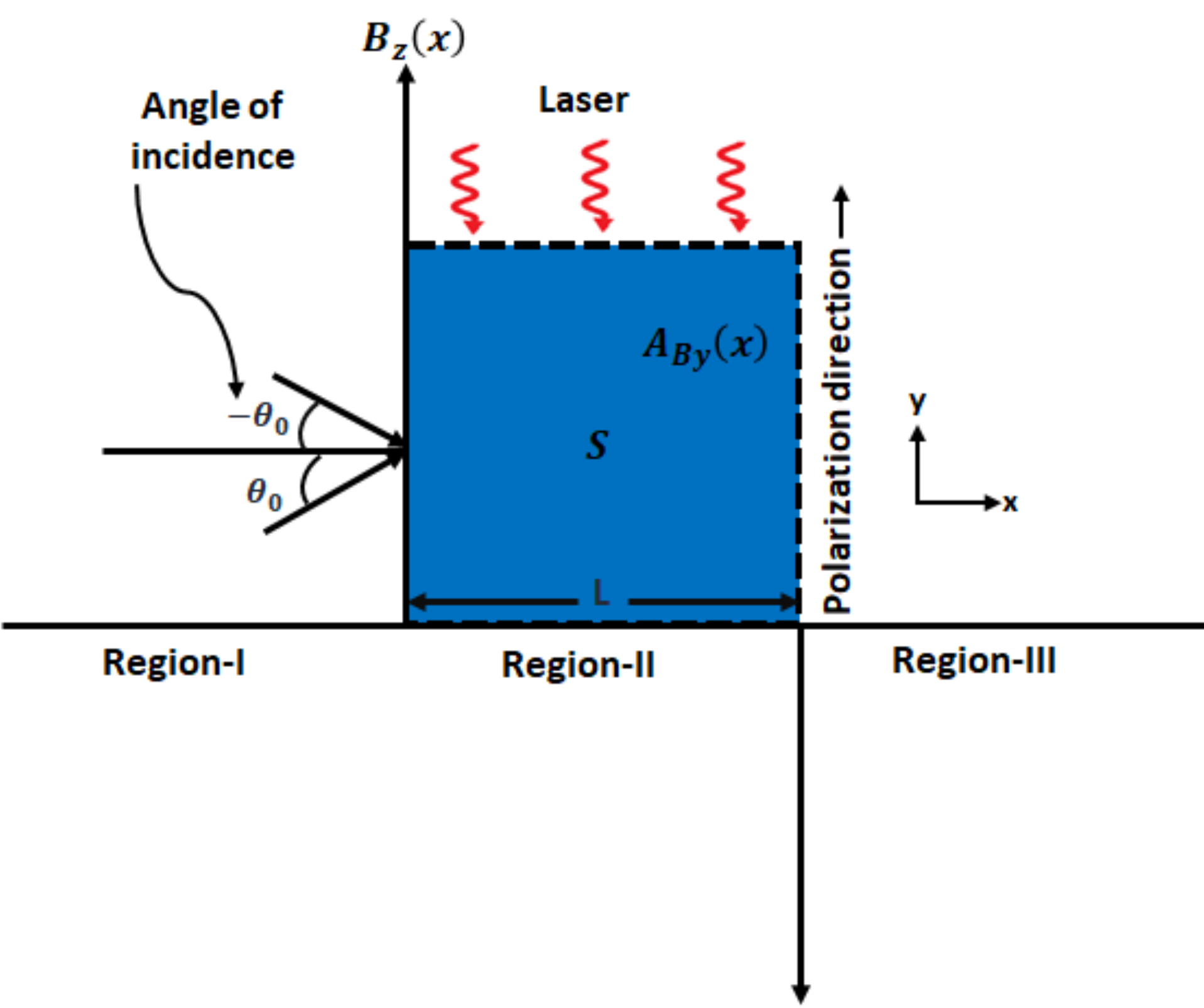}}
    \caption{{(color online) {(a)}: Normal/strained/normal graphene illuminated by a linearly polarized laser light in the strained region and subjected to a magnetic field is depicted schematically. 
    	{(b)}:  The model magnetic field ({delta-function}) {pattern} with the corresponding magnetic vector potential (dashed line)}.}\label{db.5}
    \end{figure}

We consider a graphene-based system with three regions in the $xy$ plane denoted by $j$ = I, II, and III. As shown in Fig. Fig.  \ref{db.5}a, the central region is illuminated by an electromagnetic wave that is linearly polarized along the y-axis and subjected to a magnetic field $B$ and a uniaxial strain $S$ acting in either the armchair or zigzag directions. Regarding the left and right regions, they are chosen to be pure graphene.
Before proceeding, it should be noted that in the presence of a laser field, incident electrons of energy $\varepsilon$ come from one side of the barrier at an angle $\theta_0$ and exit with the energy $\varepsilon+l\tilde\omega$ ($l = 0, \pm 1, \pm 2, \cdots $), resulting in $\theta_l$ and $\pi-\theta_l$. 
To simplify our problem, we represent a step-like magnetic potential barrier as sets of delta-function \cite{AA,bb} separated by a distance $L$, equal in length but opposite in orientation, as shown in \ref{db.5}b.

 Our system can be described by the following Hamiltonian:
\begin{equation}\label{ham2}
\mathcal{H}=v_{x}(S) {{{\sigma_x}}} \left(p_x+
    \frac{|e|}{c}\left({\color{black}{A}_{{B}x}}+{\color{black}{A}_{Lx}}\right)\right)+v_{y}(S)
    {{\color{black}{\sigma_y}}} \left(p_y+
    \frac{|e|}{c}\left({\color{black}{A}_{{B}y}}+{\color{black}{A}_{Ly}}\right)\right)
\end{equation}
where 
the canonical momentum is represented by $\boldsymbol{p}=\left(p_{x},p_{y}\right)$ and  the Pauli matrices by $\sigma_i $.
 The effective Fermi velocity components $v_x(S)$ and $v_y(S)$ are affected differently by applying a mechanical constraint on the graphene strip and are expressed as follows \cite{Sz,Yan2,Wong}
\begin{eqnarray}\label{xtyr}
    v_{x}(S)=
    \frac{\sqrt{3}c_{0}}{\hbar}\left(1+{\Lambda}_{x} S\right)
    \sqrt{t^{2}_{1}-\frac{t^{2}_{3}}{4}}, \qquad  v_{y}(S)=%\frac{1}{\hbar}
    \frac{{3}c_{0}}{2\hbar}\left(1+{\Lambda}_{y}S\right)t_{3}
\end{eqnarray}
with the modified hopping energy
\begin{align}
t_{i}=t_{0}e^{-3.37\left({|\xi_{i}|}/{c_{0}}-1\right)}	
\end{align}
where $t_{0}=2.7$ \text{eV}  is the transfer energy and $c_{0}=0.165$ \text{nm}  \cite{Sz}  is the distance between two atoms linked by a covalent bond in undeformed configurations. The displacement  of deformed graphene are given by
\begin{eqnarray}\label{xrr}
    |\xi_{1}|=|\xi_{2}|=c_{0}\left(1+\dfrac{3}{4}\Lambda_{x}S+\dfrac{1}{4}\Lambda_{y}S\right),\qquad |\xi_{3}|=c_{0}\left(1+\Lambda_{y}S\right).
\end{eqnarray}
The poisson's ratio is $({\Lambda}_{x}=-\sigma, {\Lambda}_{y}=1)$  for armchair direction and $({\Lambda}_{x}=1, {\Lambda}_{y}=-\sigma)$ for zigzag direction with $\sigma=0.165$. We assume that $v_x(S)$ and $v_y(S)$ are defined within the strip $0 \leq x \leq L$, whereas $v_x (S=0) $ and $v_y(S=0)\rightarrow v_{F}$ are defined elsewhere in this study. Alternatively, as $B_z(x)=B\left [\delta(x)-\delta(x-L)\right] $, an inhomogeneous magnetic field perpendicular to the graphene layer along the $z$-axis is chosen, and 
the separation between the two functions is represented by $L$. As shown in Fig. \ref{db.5}b, $B_z(x)$ is independent of the longitudinal transport $y$-direction and produces step-like profiles of \textbf{A}$_{B} $
	\begin{equation}\label{g2n}
		{A}_{By}(x)={Bl_{B}}\left[\Theta(x)-\Theta(x-L)\right]
\end{equation}
The typical magnetic length scale is $l_{B}=\sqrt{\dfrac{\hbar c}{|e|B}}$, and the step function is $\Theta$. The laser field is designated by 
{$\boldsymbol{A}_{L}=\frac{cF}{\omega}\left(0,\cos\omega t\right)$} where $F$ is the amplitude of electric field and $\omega$ is the wave frequency.
To go further, we will calculate  the energy spectrum and the eigenspinors of the Dirac equation.

\section{Spectral solutions}		
To find the solutions of the energy spectrum, we solve the Dirac equation for the spinor
\begin{equation}\label{hj1}
\Phi(x,y,t)=e^{-i\frac{v_F}{l_{B}}\varepsilon {t}}\phi(x, y, t)=\left[\Phi^{\text{A}}(x,y,t),\Phi^{\text{B}}(x, y, t)\right]^{\text{T}}
\end{equation} 
where the Floquet energy is $\varepsilon= \frac{E}{E_0}$, and ${E_0}= \frac{\hbar v_{F}}{l_{B}}$.  The periodic function $\phi(x,y, t) $ confirms that $\phi_j(x,y, t+2\pi/\omega) =\phi_j(x,y, t) $ is a Fourier expansion 
\begin{align}\label{gg}
\phi(x,y,t)=\varphi(x,y) \sum^{+\infty}_{m=-\infty}J_{m}\left(\dfrac{\tilde F \tilde v_{y}(S)}{\tilde\omega^{2}}\right) e^{-i\frac{v_F}{l_{B}}m{\tilde\omega}t}
\end{align}
with $\tilde{F}=\frac{F |e|l_{B}}{E_{0}}$, ${\tilde{\omega}=\frac{\omega l_{B}}{v_{F}}}$,
$\tilde{v}_{y}(S)=\frac{v_{y}(S)}{v_F}$ and 
the first kind of Bessel function is $J_{m}$. We write the wave function $\varphi (x,y) =e^{ik_{y}y}\varphi(x)$ in separable form by considering the conservation of the transverse $p_y$.

We consider the Dirac equation in the following form for  region I ($x<0$),
\begin{align}\label{eq9}
        l_B\begin{pmatrix}
        0 & -i\partial_{x}-ik_{y} \\
        -i\partial_{x}+ik_{y} & 0 \\
    \end{pmatrix}%
    \begin{pmatrix}
    {\varphi^{\text{A}}_{\textbf{I}}(x)} \\
    {\varphi^{\text{B}}_{\textbf{I}}(x)}\\
    \end{pmatrix}%
    =\varepsilon
    \begin{pmatrix}
            {\varphi^{\text{A}}_{\textbf{I}}(x)}  \\
            {\varphi^{\text{B}}_{\textbf{I}}(x)}  \\
    \end{pmatrix}%
\end{align}
and then obtain
\begin{eqnarray}
	&& \left(-i\partial_{x}-ik_{y} \right)\varphi^{\text{B}}_{\textbf{I}}(x) =\dfrac{{\varepsilon}}{{l_{B}}}
	%\frac{{E}}{\hbar v_{F}} 	
	\varphi^{\text{A}}_{\textbf{I}}(x) \\
	&& \left(-i\partial_{x}+ik_{y}\right)	\varphi^{\text{A}}_{\textbf{I}}(x) =\dfrac{{\varepsilon}}{{l_{B}}}
	%\frac{{E}}{\hbar v_{F}}
	\varphi^{\text{B}}_{\textbf{I}}(x).
\end{eqnarray}
This may be utilized to display finally $\Phi_{\textbf{inc}}(x,y,t)$ as
\begin{equation}
	{\Phi_{\textbf{inc}}(x,y,t)}=
	\begin{pmatrix}
		1 \\
		{{{z}_{0}}}\end{pmatrix}e^{ik^{0}_{x}x+ik_{y}y}e^{-i\frac{v_F}{l_{B}} \varepsilon t}
\end{equation}
where the complex number ${z}_{0}$ and the incident angle $\theta_{0}$ are given by
\begin{equation}
{{{z}_{0}}}=
    s_{0} e^{\textbf{\emph{i}} {\theta_{0}}}, \qquad \theta_{0}=\tan^{-1}\left(\dfrac{k_{y}}{k^{0}_{x}}\right)
\end{equation}
with $s_{0}=\mbox{sgn}(\varepsilon)$. It is shown that the components of reflected and transmitted eigenspinors exist at all energies ${\varepsilon}+ l{\tilde\omega} $ $(l=0,\pm 1,\cdots)$ {\bb\cite{A1,A11}}. As a result, the eigenspinors $\Phi_{{\textbf{ref}}}(x,y,t)$  are discovered
\begin{align}
	{\Phi_{\textbf{ref}}(x,y, t)}=\sum^{+\infty}_{l,m=-\infty} {r_{l}}
	\begin{pmatrix}
		1 \\
		-\dfrac{1}{{{z}_{l}}}\end{pmatrix}e^{-ik^{l}_{x} x +ik_{y}
		y} J_{m-l}{{\left(\frac{\tilde F}{\tilde\omega^{2}}\right)}}\ e^{-i\frac{v_F}{l_{B}}\left({\varepsilon}+m{\tilde\omega}\right) t}
\end{align}
where the complex number and the angle of the reflected electrons are defined by
\begin{equation}
{{z}_{l}}=
s_{l} e^{\textbf{\emph{i}}{\theta_{l}}}, \qquad {\theta_{l}}=\tan^{-1}\left(\dfrac{k_{y}}{k^{l}_{x}}\right)
\end{equation}
with $r_ l$ is the reflection amplitude and $s_{l}=\mbox{sgn}({\varepsilon}+l\tilde{\omega})$ is the sign function.
The corresponding eigenvalues read as
\begin{equation}\label{energy1}
{\varepsilon}+l\tilde\omega=s_{l}l_B\sqrt{(k^{l}_{x})^{2}+ k^{2}_{y}}.
\end{equation}
From \eqref{energy1}, we obtain the  wave vector
\begin{equation}
k^{l}_{x}=\frac{s_{l}}{l_B}\sqrt{{\left({\varepsilon}+l\tilde{\omega}\right)^{2}}-(k_{y}l_{B})^{2}}.
\end{equation}
We can write $J_{m-l}\left(0\right)={\delta_{ml}}$ by ignoring the laser field, i.e., $F=0$. 
Combining them all together to get the eigenspinors in region I ($x<0$) as
\begin{equation}
{\Phi_{\text{I}}(x,y,t)}=e^{ik_{y}y}\sum^{+\infty}_{l,m=-\infty}\left[\delta_{l0}
\begin{pmatrix}
        1 \\
{{{z}_{l}}}\end{pmatrix}e^{ik^{l}_{x}x}+{r_{l}}
\begin{pmatrix}
1 \\
-\dfrac{1}{{{z}_{l}}}\end{pmatrix}e^{-ik^{l}_{x} x}\right]\delta_{ml}\  e^{-i\frac{v_F}{l_B}\left({\varepsilon}+m{\tilde\omega}\right)t}.
\end{equation}
In the case of region III ($x>L$), we express the eigenspinors  {$\Phi_{\sf{III}}(x,y,t)$} as in region I
    \begin{equation}
        {\Phi_{\text{III}}(x,y,t)}=e^{ik_{y}y}\sum^{+\infty}_{l,m=-\infty}\left[t_{l}\left(
        \begin{array}{c}
            1 \\
            {{z}_{l}}\end{array}\right)e^{ik^{l}_{x}x
        }+\beta_{l}\left(
        \begin{array}{c}
            1 \\
            -\dfrac{1}{{{z}_{l}}}\end{array}\right)e^{-ik^{l}_{x}x}\right]\delta_{ml}\
        e^{-i\frac{v_F}{l_B}\left({\varepsilon}+m{\tilde\omega}\right)t}.
    \end{equation}
with $t_l$ being the transmission amplitude and $\{\beta_{l}\}$ being the null vector. 

The eigenvalues equation for  region II ($x\leq0\leq L$)  is written as 
\begin{align}\label{hy}
        \begin{pmatrix}
            \varepsilon+l\tilde\omega   &  i\left(\tilde v_{x}(S)\partial_{x}{l_B}+\tilde v_{y}(S)\left[k_{y}{l_B}+{1}\right]+l\tilde\omega\right) \\
        i\left(\tilde v_{x}(S)\partial_{x}{l_B}-\tilde v_{y}(S)\left[k_{y}{l_B}+{1}\right]-l\tilde \omega\right)&  \varepsilon+l\tilde\omega
        \end{pmatrix}\begin{pmatrix}
            \varphi^{\text{A},l}_{{\textbf{II}}}(x)\\
            \varphi^{\text{B},l}_{{\textbf{II}}}(x)
        \end{pmatrix}=\begin{pmatrix} 0\\0
        \end{pmatrix}
\end{align}
    and the corresponding eigenvalues are   
    \begin{align}\label{Aqf}
        {\varepsilon}+l\tilde\omega&=s'_l\sqrt{{{\tilde{v}^{2}_{x}(S)(q^{l}_{x}l_{B})^{2}+\left({{\tilde{v}_{y}(S)}\left[k_{y}l_{B}+1\right]+l\tilde\omega}\right)^{2}}}}
   \end{align}
where
$\tilde{v}_{x}(S)= \frac{v_{x}(S)}{v_F}$ and $s'_l=\text{sgn}\left(\varepsilon+l\tilde\omega\right)$. The wave vector $q^{l}_{x}$  can be found in (\ref{Aqf})  as 
   {\begin{align}\label{Ahq}
 q^{l}_{x}=\frac{s'_l}{l_B}\sqrt{\dfrac{\left(\varepsilon+l\tilde\omega\right)^{2}}{\tilde{v}^{2}_{x}(S)} -\left(\dfrac{{\tilde{v}_{y}(S)}\left[k_{y}l_{B}+1\right]+l\tilde\omega}{\tilde{v}_{x}(S)}\right)^{2}}.
    \end{align}}
At this point, we note that the laser light dresses the wave vector's $x$-component, and the shift in $k_y$ is due to the presence of the vector potential {\bb\cite{str1,qqa,dr1}}, which is not the case for a scalar potential that depends on time \cite{A1,A2,A6,A71,A7}.  As a result, the appropriate eigenspinors are identified as 
\begin{align}
\varphi^{l}_{\text{II}}(x,y)&=e^{ik_{y}y}\sum^{
	+\infty}_{l=-\infty} \left[c_{1,l}\begin{pmatrix}
	1\\
	{{z'_{l}}}
\end{pmatrix}e^{iq^{l}_{x}x}+c_{2,l} \begin{pmatrix}
	1\\
	-\dfrac{1}{{{z'_{l}}}}
\end{pmatrix}e^{-iq^{l}_{x}x}\right].
\end{align}
Finally, the eigenspinors of  region II ($0\leq x\leq L$) are calculated
\begin{align}
	\Phi_{\text{II}}(x,y,t)&=e^{ik_{y}y}\sum^{
		+\infty}_{l,m=-\infty} \left[c_{1,l}\begin{pmatrix}
		1\\
		{{{z'}_{l}}}
	\end{pmatrix}e^{iq^{l}_{x}x}+c_{2,l} \begin{pmatrix}
		1\\
		-\dfrac{1}{{{z'_{l}}}}
	\end{pmatrix}e^{-iq^{l}_{x}x}\right]J_{m-l}{\left(\frac{\tilde{F}\tilde{v}_{y}(S)}{\tilde{\omega}^{2}}\right)}e^{-i\frac{v_F}{l_B}\left({\varepsilon}+m{\tilde\omega}\right)t}
\end{align}
where $c_{1,l}$, $c_{2,l}$ are two constants and we have set {the complex number $z'_{l}$ and the barrier's inside angle $\theta'_{l}$ as}
\begin{align}
        {{{z'_{l}}}}=
        {s'_l}e^{i\theta'_{l}}, \qquad {\theta'_{l}=\tan^{-1}\left(\dfrac{ {{\tilde{v}_{y}(S)}}\left[k_{y}l_B+1\right]+l\tilde\omega}{{\tilde{v}_{x}(S)} q^{l}_{x}l_B}\right)}.
\end{align}
The preceding results show that the Floquet eigenvalues and solutions of region II are highly dependent on frequency $\omega $, magnetic field $B$, uniaxial deformation $S$, laser light amplitude $F$, and wave vector components.  
{We recover the results of \cite{Xu5} when the effective velocities $v_x(S)$ and $v_y(S)$ fall to Fermi values for zero strain, $S=0$.}
In the forthcoming analysis, we will study the impact of armchair and zigzag strain on transmission probability via delta-function magnetic barrier.

\section{Transmission through laser strained magnetic barrier}\label{Tran}

In our deformed graphene system, we will look at the transmission probability via a laser-assisted magnetic barrier.
For this, we use the continuity of  wave functions at two interfaces $(x=0,x={L})$ 
\begin{eqnarray}
    && {\Phi_{\text{I}}(0,y,t)}={\Phi_{\text{II}}(0,y,t)} \label{psix0}\\
    && {\Phi_{\text{II}}({L},y,t)}={\Phi_{\text{III}}({L},y,t)} \label{psixd}.
\end{eqnarray}
Using the orthogonality of $\{e^{im{\omega} t}\}$, we write at interface $x=0$
\begin{align}
  &  \delta_{m0}+r_{m} =\sum^{+\infty}_{l=-\infty}
    \left[c_{1,l}+c_{2,l}\right]
    J_{m-l}{{\left(\frac{\tilde{F}\tilde{v}_{y}(S)}{\tilde{\omega}^{2}}\right)}} \label{eqx01}\\
  & \delta_{m0}{{z_{m}}}-r_{m}\frac{1}{{{z_{m}}}} =
    \sum^{+\infty}_{l=-\infty}{\left[c_{1,l}{z'_{l}}-c_{2,l}\frac{1}{{z'_{l}}}\right]}
    J_{m-l}{{\left(\frac{\tilde{F}\tilde{v}_{y}(S)}{\tilde{\omega}^{2}}\right)}} \label{eqx02}
\end{align}
and for $x=L$, we have
\begin{align}
    %\left\{\begin{array}{c}
  &  t_{m}{e^{ik^{m}_{x}l_B{\frac{L}{l_B}}}}+
    {\beta_{m}}{e^{-ik^{m}_{x}l_B{\frac{L}{l_B}}}} =\sum^{+\infty}_{l=-\infty} \left[c_{1,l}{e^{iq^{l}_{x}l_B\frac{L}{l_B}}}+c_{2,l}{e^{-iq^{l}_{x}l_B\frac{L}{l_B}}}\right]J_{m-l} {{\left(\frac{\tilde{F}\tilde{v}_{y}(S)}{\tilde{\omega}^{2}}\right)}} \label{eqxd1} \\
  & t_{m}{{z_{m}}}{e^{ik^{m}_{x}l_B{\frac{L}{l_B}}}}-{\beta_{m}}\frac{1}{{z_{m}}}{e^{-ik^{m}_{x}l_B{\frac{L}{l_B}}}} 
    =\sum^{+\infty}_{l=-\infty}\left[c_{1,l}{z'_{l}}{e^{iq^{l}_{x}l_B\frac{L}{l_B}}}-c_{2,l}\frac{1}{{z'_{l}}}{e^{-iq^{l}_{x}l_B\frac{L}{l_B}}}\right]J_{m-l} {{\left(\frac{\tilde{F}\tilde{v}_{y}(S)}{\tilde{\omega}^{2}}\right)}}. \label{eqxd2}
\end{align}
After calculation, we can represent these boundary conditions in the transfer matrix formalism
\begin{align}\label{ar1}
    \begin{pmatrix}
    {\Xi_{1}}\\
    {\Xi_{1}^{'}}
    \end{pmatrix}=\begin{pmatrix}
    {\mathbb{W}_{11}}   &{\mathbb{W}_{12}}\\
    {\mathbb{W}_{21}}&{\mathbb{W}_{22}}
\end{pmatrix}\begin{pmatrix}
\Xi_{2}\\
\Xi_{2}^{'}
\end{pmatrix}={\mathbb W}\begin{pmatrix}
    \Xi_{2}\\
    \Xi_{2}^{'}
    \end{pmatrix}
\end{align}
which can be expressed as
\begin{eqnarray}
{\mathbb W}={\mathbb W^{-1}_{1}}(0)\cdot{\mathbb W_{2}}(0)\cdot{\mathbb W^{-1}_{2}}(L)\cdot{\mathbb W_{1}}(0)\cdot{\mathbb W_{3}}(L)
\end{eqnarray}
where {we have defined the following matrices}
    \begin{align}
&{\mathbb W_{1}}(0)=
    \begin{pmatrix}
        {\mathbb I}& {\mathbb I} \\
        {{\mathbb C^{+}}} &{{\mathbb C^{-}}} \\
    \end{pmatrix}\\
 &{\mathbb W_{2}}(x)=
    \begin{pmatrix}
        {{\mathbb D^{+}(x)}} & {{\mathbb D^{-}(x)}} \\
        {{\mathbb P^{+}(x)}} & {{\mathbb P^{-}(x)}}
    \end{pmatrix}\\
& {\mathbb W_{3}}(L)=
\begin{pmatrix}
{{\mathbb Q^{+}}}& {\mathbb O} \\
{{\mathbb O}} &{{\mathbb Q^{-}}} \\
\end{pmatrix}
\end{align}
with the associated matrix elements
\begin{align} \label{eqn1}
    &\left({{\mathbb
            C^{\pm}}}\right)_{ml}=\pm\left({z_{{m}}}\right)^{\pm
        1}\delta_{ml}\\
  &  \left({{\mathbb
            D^{\pm}(x)}}\right)_{ml}=e^{\pm iq^{l}_{x}{l_{B}}\frac{x}{{l_{B}}}}J_{m-l}{{\left(\frac{\tilde{F}\tilde{v}_{y}(S)}{\tilde{\omega}^{2}}\right)}} \\ &\left({{\mathbb P^{\pm}(x)}}\right)_{ml}=\pm ({z'_{l}})^{\pm 1}e^{\pm iq^{l}_{x}{l_{B}}\frac{x}{{l_{B}}}}
    J_{m-l}{{\left(\frac{\tilde{F}\tilde{v}_{y}(S)}{\tilde{\omega}^{2}}\right)}}\\
    &{\left({\mathbb  Q^{\pm}}\right)_{ml}=e^{\pm ik^{m}_{x}{l_{B}}\frac{L}{{l_{B}}}}\delta_{ml}}. 
\end{align}
The unit and null matrices are denoted by ${\mathbb O}$ and ${\mathbb I}$, respectively. Then, ${\Xi_{1}}=\{\delta_{0l}\}$, ${\Xi_{1}^{'}}=\{r_{l}\}$,  $\Xi_{2}=\{t_{l}\}$ and $\Xi_{2}^{'}=\{\beta_{l}\}=0$ with $x=(0,L)$. Based on the foregoing results, we end up with
\begin{align}
\Xi_{2}={\mathbb{W}_{11}^{-1} \cdot\Xi_{1}}.
\end{align}
 The infinite series for $T_l$ can be broken down into a finite number of terms beginning with $-N$ and ending with $N$ {\cite{ds,A6,A71,A7}}, {where $N>\frac{\tilde{F}\tilde{v}_{y}(S)}{\tilde{\omega}^{2}}$}. This yields the outcome 
\begin{equation}
    t_{-N+k}={\mathbb{W}'}\left[k+1, N+1\right]\label{eqt}
\end{equation}
with $k=0, 1,\cdots, 2N$ and {${\mathbb W^{'}}$} is the inverse matrix ${\mathbb{W}_{11}^{-1}}$.

To calculate the transmission probability $T_{l}$ corresponding to our system, we introduce the transmitted $J_{\sf \textbf{tran},l}$ and the reflected $J_{\sf \textbf{inc}, 0}$ currents. These are described by the relation \cite{A6}
\begin{align}\label{ma3}
T_{l}=\left|\dfrac{J_{\sf \textbf{tra},l}}{J_{\sf \textbf{inc},0}}\right|.
\end{align}
To evaluate (\ref{ma3}), we introduce
\begin{align}\label{ma4a}
    J=|e| v_{F}\Phi^{\dagger}(x,y,t)\sigma_{x}\Phi(x,y,t).
\end{align}
After using the energy spectrum solutions for each region and injecting it in (\ref{ma3}), we get up
\begin{align}\label{ma4}
T_{l}=\dfrac{k_{l}}{k_{0}}|t_{l}|^{2}.
\end{align}
The sum of overall channels, $l$ gives the total transmission probability through a laser-assisted strained magnetic barrier
\begin{align}
    T_{c}=\sum_{l}T_{l}.
\end{align}
We can truncate (\ref{eqt}) and keep only the terms corresponding to the central band, $l=0$, and the two first sidebands, $l=\pm 1$, in the following analysis
%.
%In the analysis to come, we can truncate (\ref{eqt}) and preserve just the terms corresponding to the central band, $l=0$ and the two first sidebands, $l=\pm 1$
\begin{equation}\label{gt}
    t_{-1}={\mathbb W^{'}}[1,2], \qquad t_{0}={\mathbb W^{'}}[2,2], \qquad t_{1}={\mathbb W^{'}}[3,2].
\end{equation}
%Therefore, we will numerically show the impact of $S$ on the manifestations of transmission probability $T_{c}$ under various physical parameters of our system.
As a result, we will numerically demonstrate the effect of $S$ on the manifestations of transmission probability $T_{c}$ for a variety of physical parameters in our system. 

\section{Results and discussions}{\label{ANA}}

\begin{figure}[H]
	\centering
	{\subfloat[]{
			\includegraphics[width=0.45\linewidth, height=0.19\textheight]{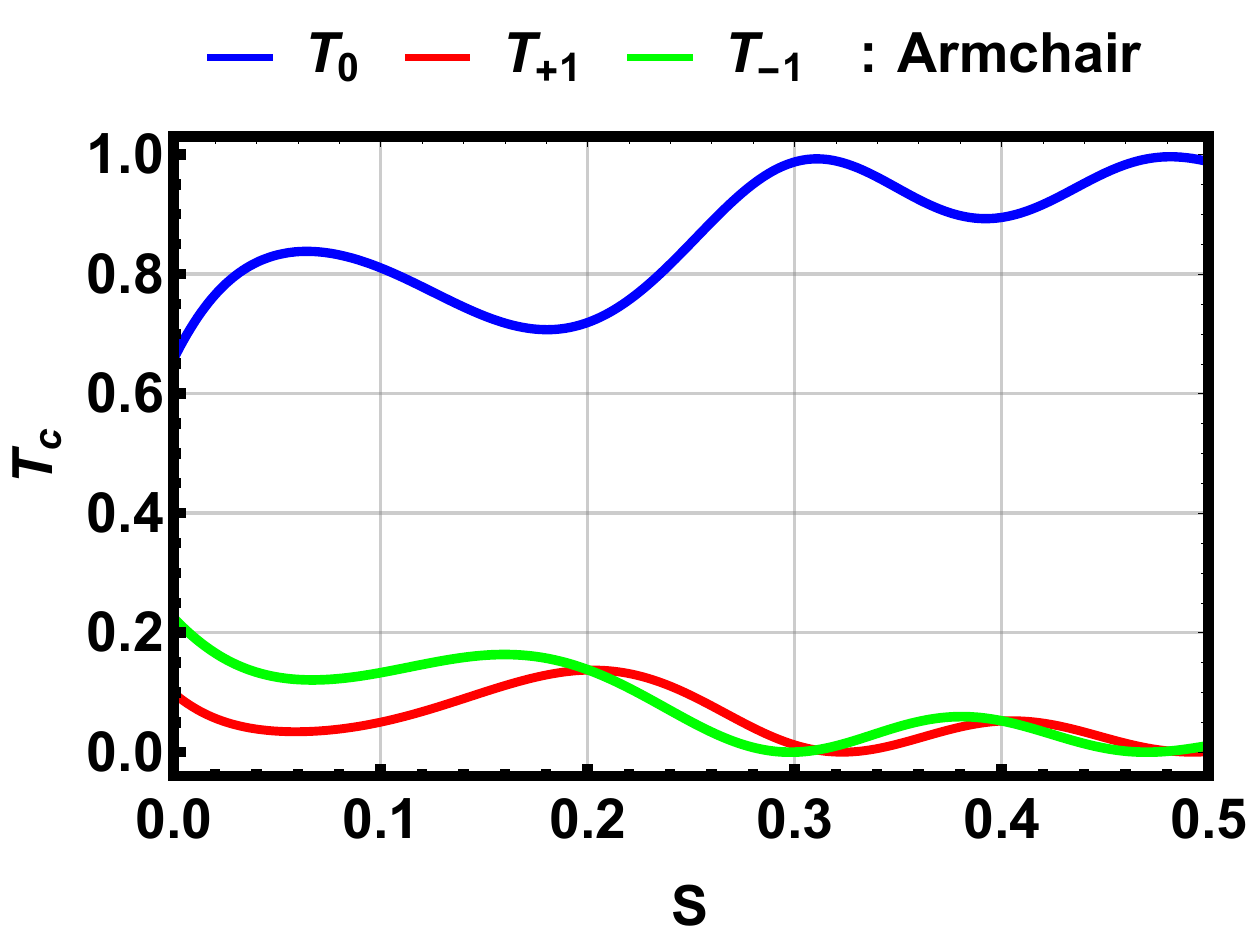}}}
	{\subfloat[]{
			\includegraphics[width=0.46\linewidth, height=0.1909\textheight]{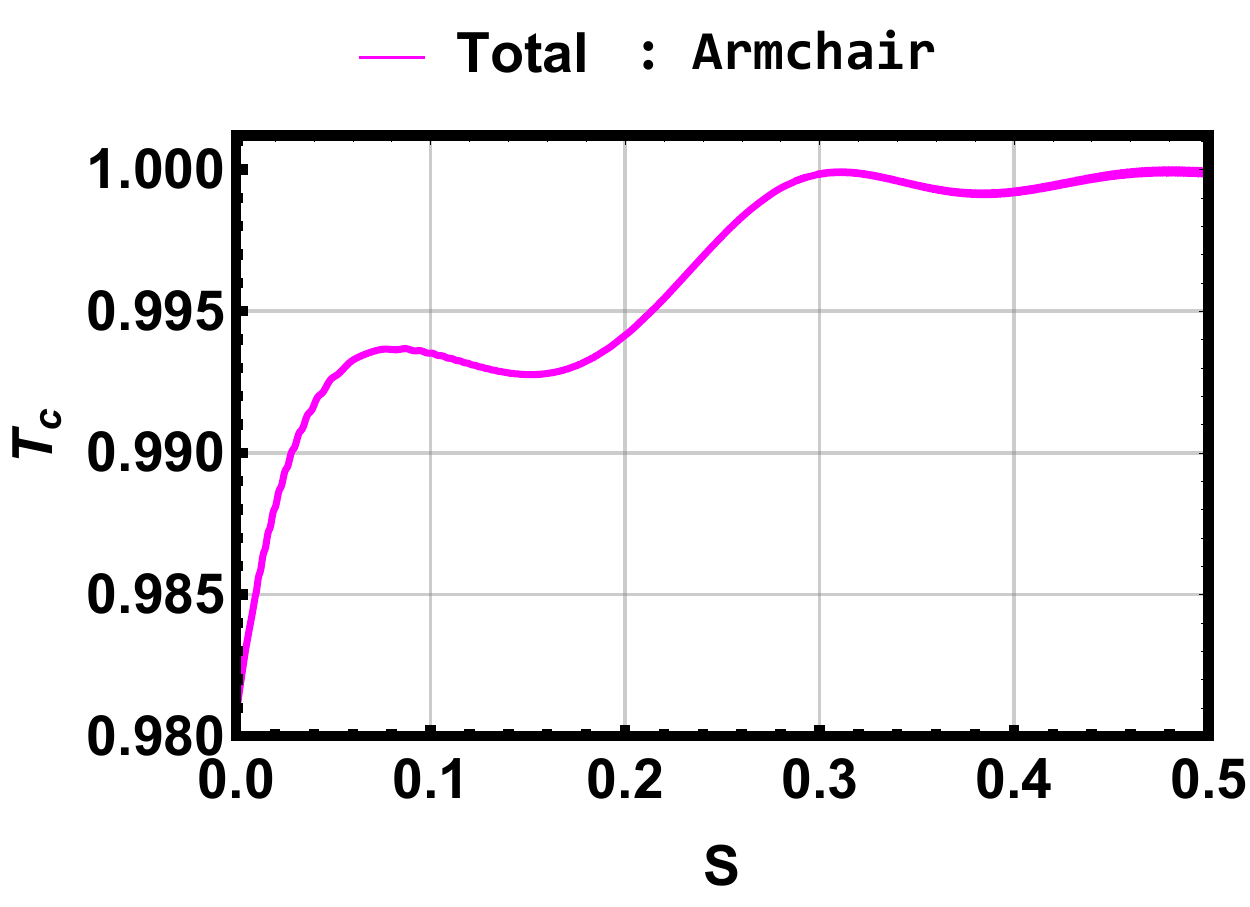}}}\\
	{\subfloat[]{
			\includegraphics[width=0.45\linewidth, height=0.19\textheight]{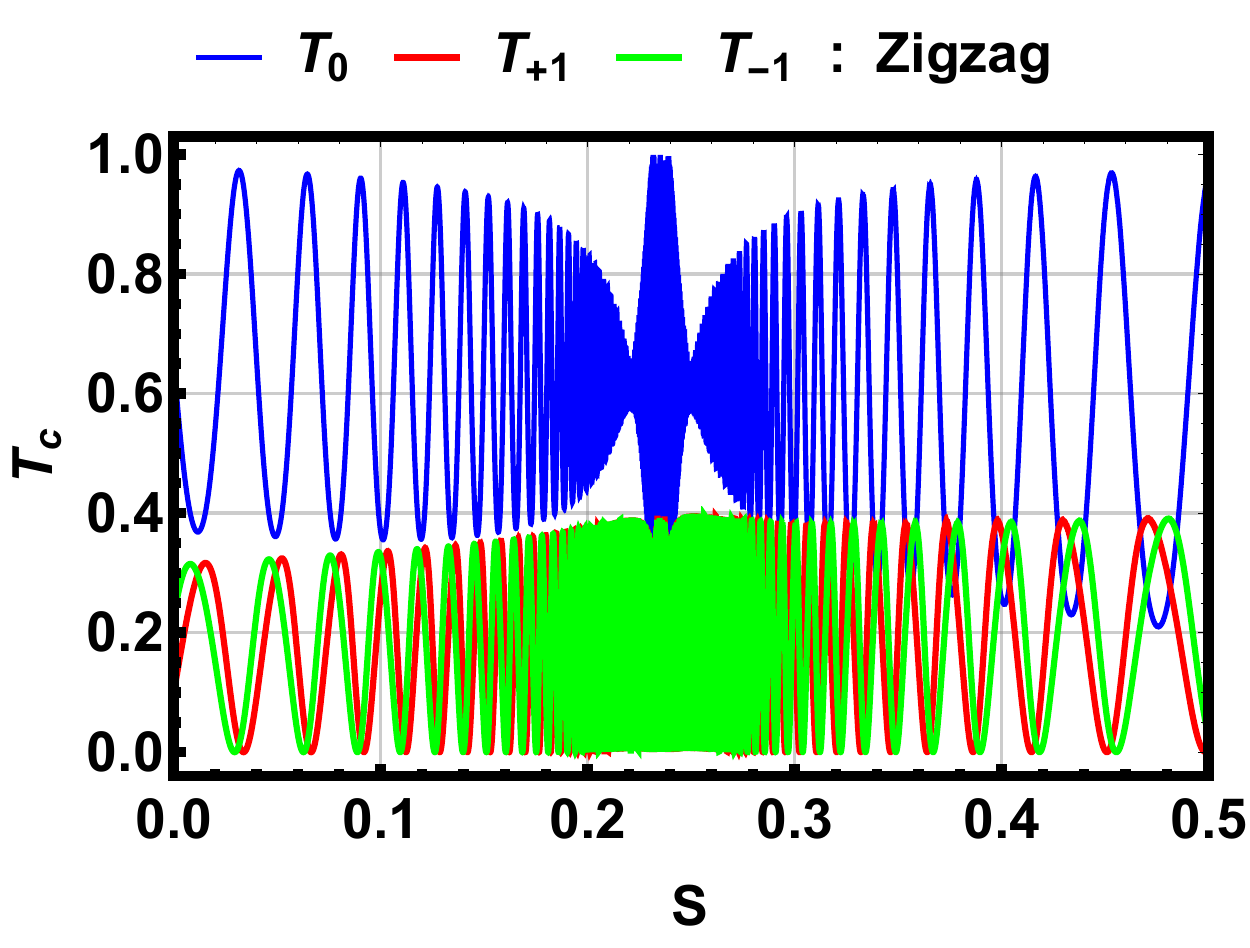}}}	
	{\subfloat[]{
			\includegraphics[width=0.46\linewidth, height=0.1909\textheight]{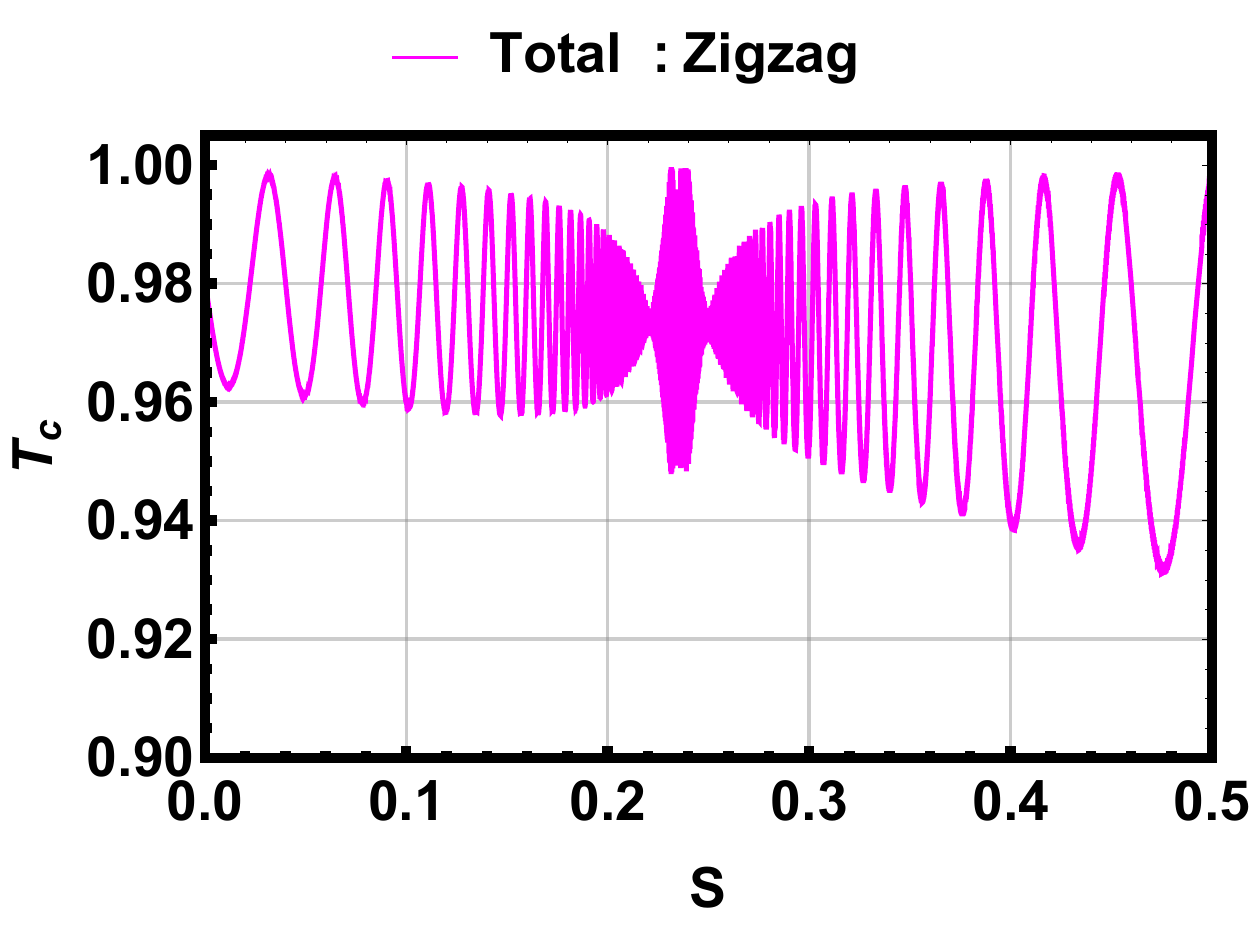}}}
	\caption{(color online) 
		Transmission probabilities  for central band $T_{0}$, first sidebands $T_{\pm1}$ and  $T_{c}$ versus the strain amplitude $S$ along armchair and zigzag directions for {$\varepsilon=75$, ${L}/{l_B}=60$, $\tilde{F}=0.658$, $\tilde{\omega}=1$ and $k_{y}l_B=1.5$.}}\label{fi}
\end{figure}

Fig. \ref{fi}
depicts the effect of strain amplitude along the armchair and zigzag directions on the central band $T_0$, first sidebands $T_{\pm1}$, and total transmission $T_{c}$, 
%$\sum_{l}T_{l}$ for $l=-1$ to $1$, 
with $\varepsilon=75$, ${L}/{l_B}=60$, $\tilde{F}=0.658$, $\tilde{\omega}=1$ and $k_{y}l_B=1.5$.
We can see from Fig. \ref{fi}a that the transmission through the central band $T_ 0 $ rises monotonically along the armchair strain direction until it approaches unity ($T_{c}\sim1$) 
whereas the transmission for the sidebands oscillates decreasingly and becomes null for larger values of $S$. 
{As shown in Fig. \ref{fi}c, the zigzag deformation has an admirable influence, as the number of oscillations for all channels increases rapidly and gradually under the condition $S\lesssim0.23$, but its maxima begin to decrease. Of course, the altered forms of $q^{l}_x$ and $\theta'_l$ in the central region are the cause of this diminishing.
We can clearly see that the transmission exhibits symmetry as $S$ increases, in contrast to the results obtained for the case of time periodic scalar potential \cite{Yan2}, where $T_c$ shows only ripples. 
}   
In Fig.  \ref{fi}(b,d)  
the total transmission will be focused on. It is observed that there is a clear change between the two cases of strain magnitudes. Speaking more generally, the total transmission shows a slight oscillatory undulation where it remains constant ($T_{c}\sim1$) at $S\gtrsim0.45$ as the strain is applied along the armchair direction. {We can say that this behavior is roughly similar to that attained in \cite{Yan2}.} In contrast, in the zigzag case, the strain has a significant effect on the oscillations of $T_c$, where their number increases dramatically, which agrees with the earlier analysis. 
Note that the behaviors of the total transmission are somewhat similar to those obtained for no photon exchange $T_{0}$, except that $T_{c}$ is shifted to the up.  Finally, it is worth noting that $S$ plays an important role in transmission and could have a significant impact on our system's tunneling features. 

\begin{figure}[ht] \centering
  	{\subfloat[]{
  			\includegraphics[width=0.45\linewidth, height=0.19\textheight]{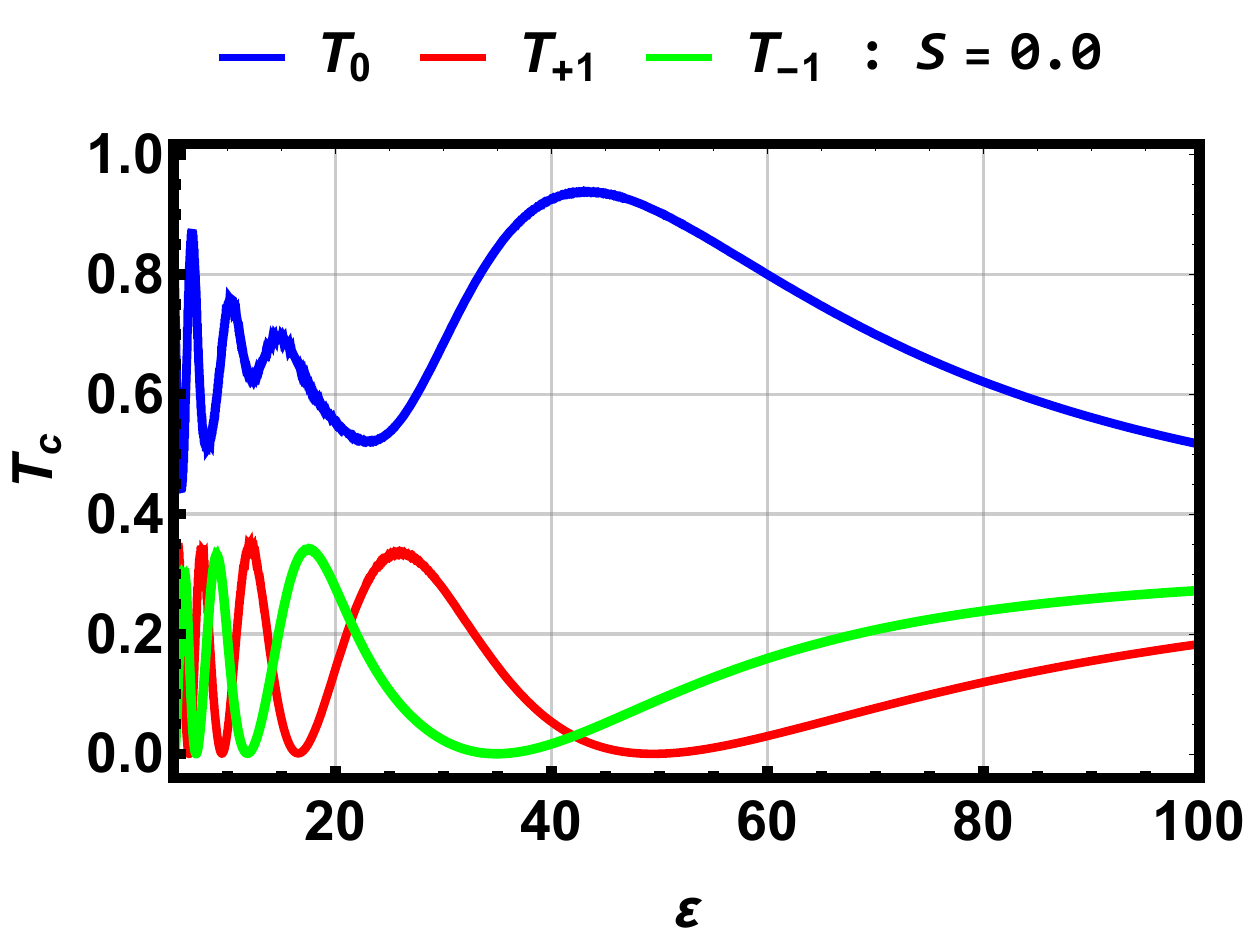}}}
  	{	\subfloat[]{
  			\includegraphics[width=0.46\linewidth, height=0.1935\textheight]{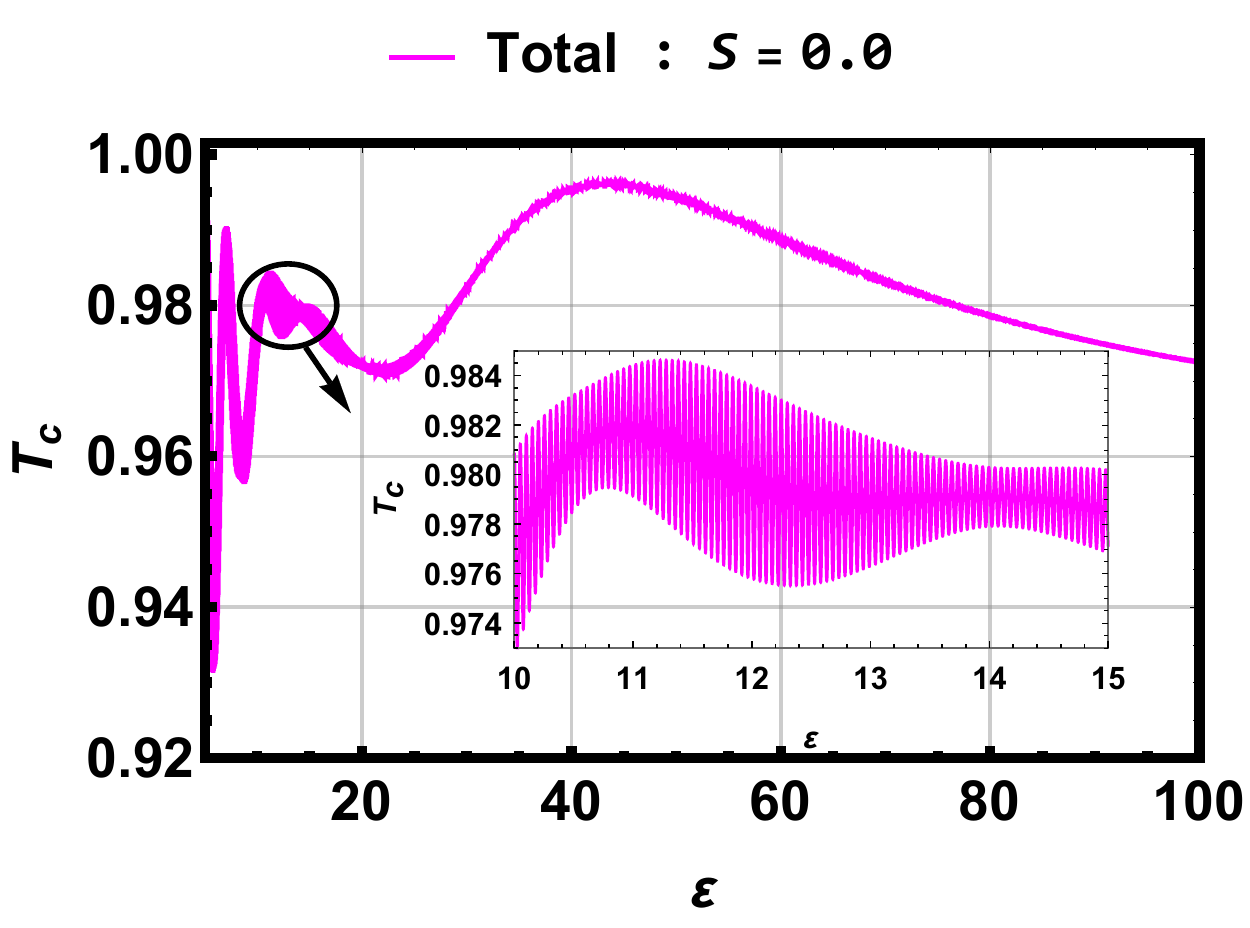}}}\\
  	{\subfloat[]{
  			\includegraphics[width=0.45\linewidth, height=0.19\textheight]{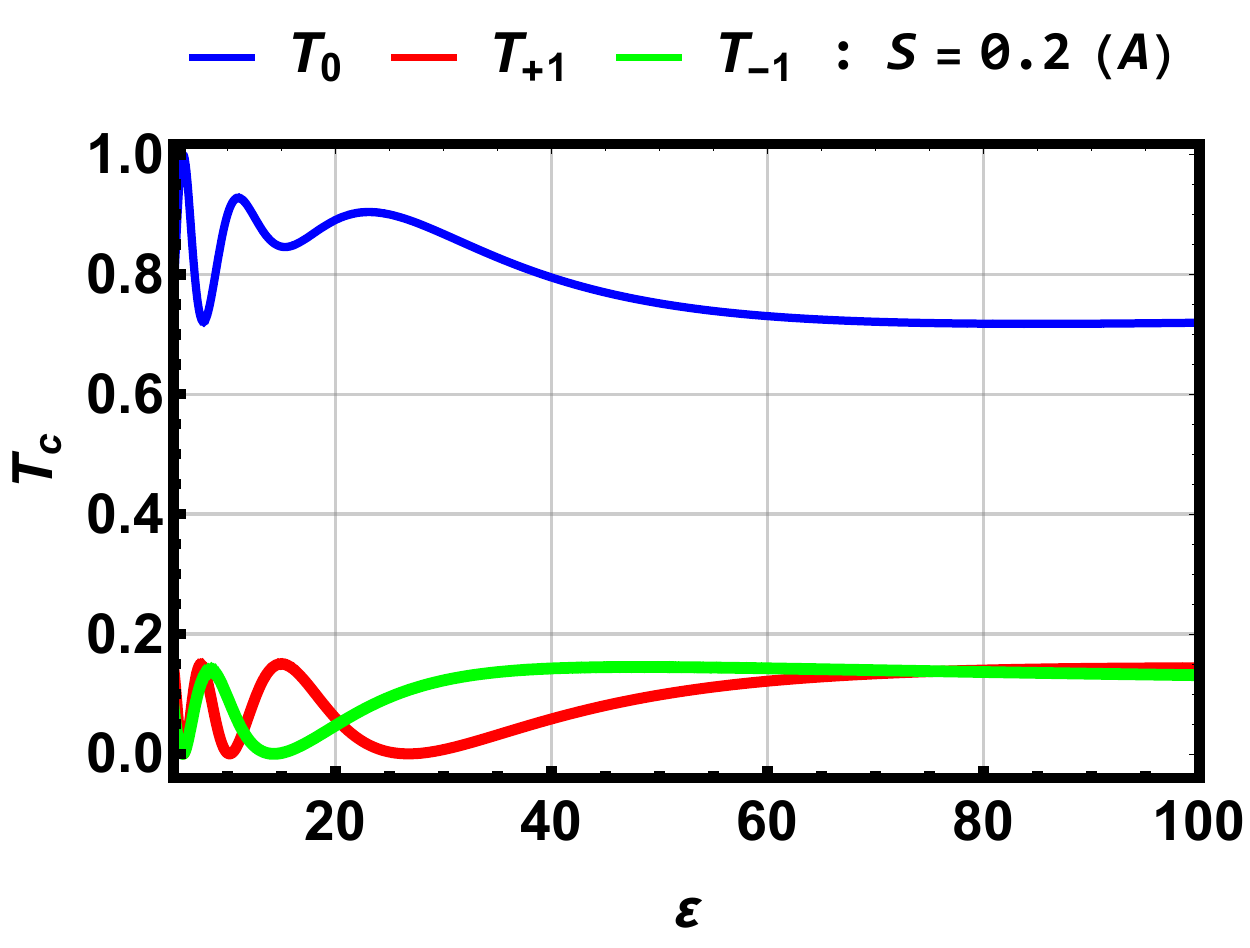}}}
  	{	\subfloat[]{
  			\includegraphics[width=0.46\linewidth, height=0.1935\textheight]{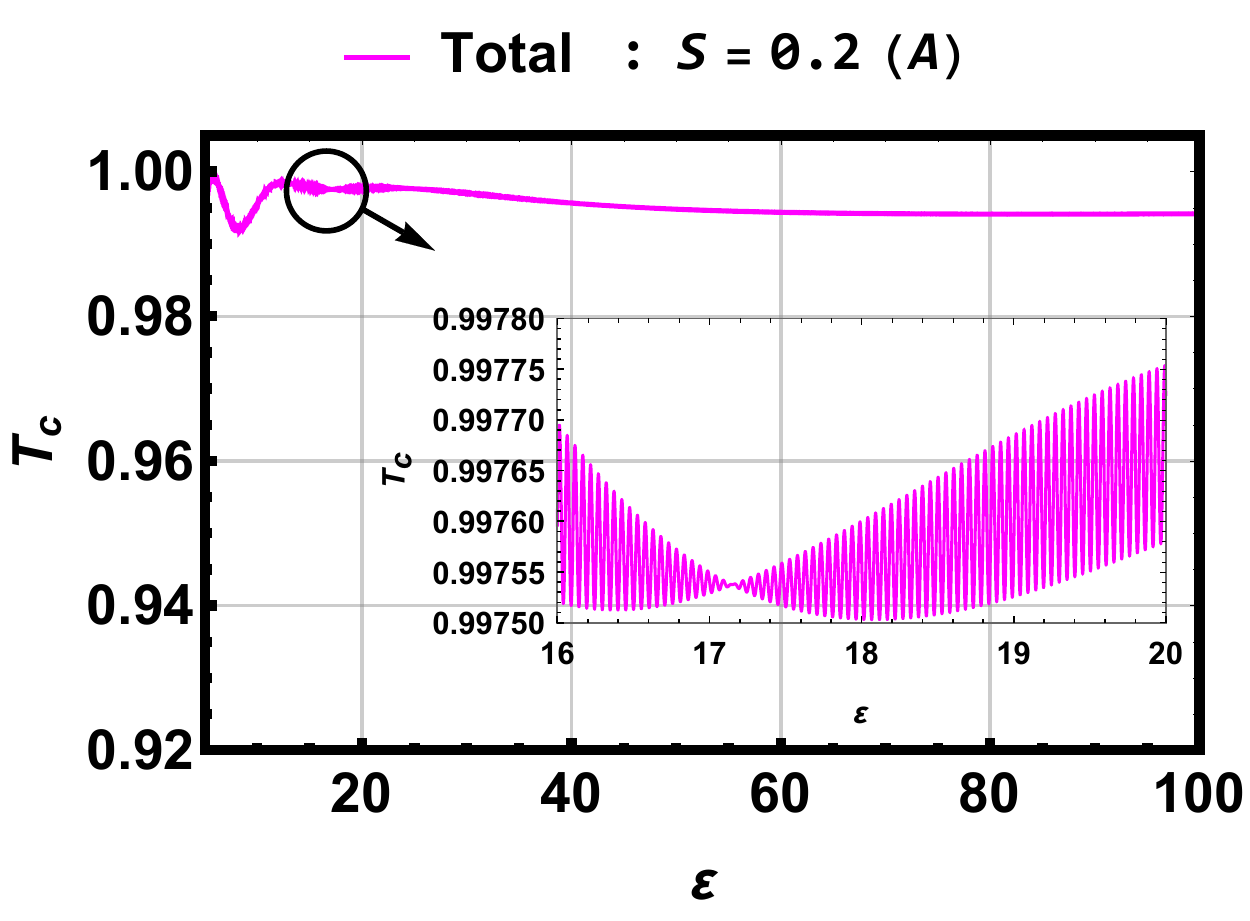}}}\\
  	{\subfloat[]{
  			\includegraphics[width=0.45\linewidth, height=0.19\textheight]{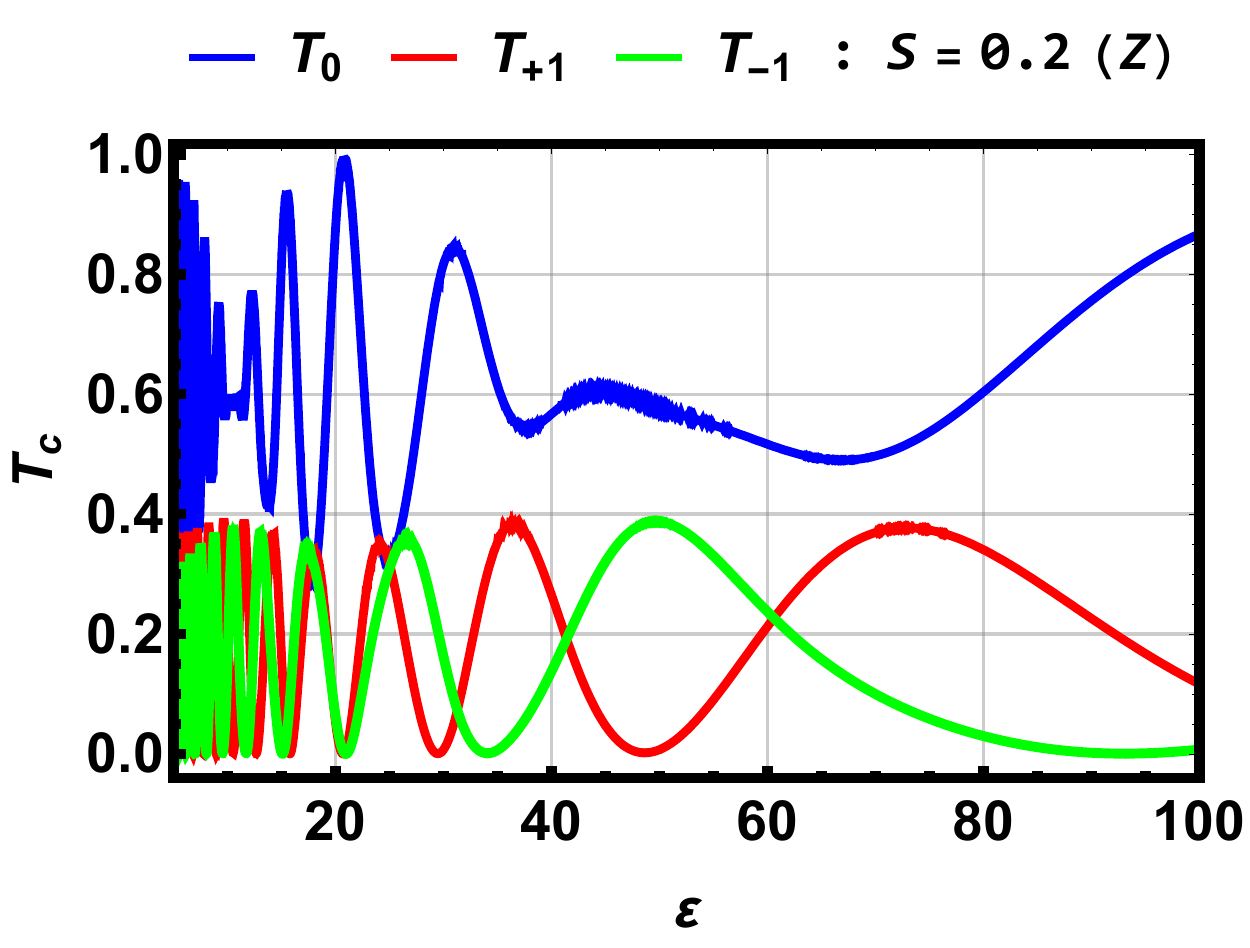}}}
  	{	\subfloat[]{
  			\includegraphics[width=0.46\linewidth, height=0.1935\textheight]{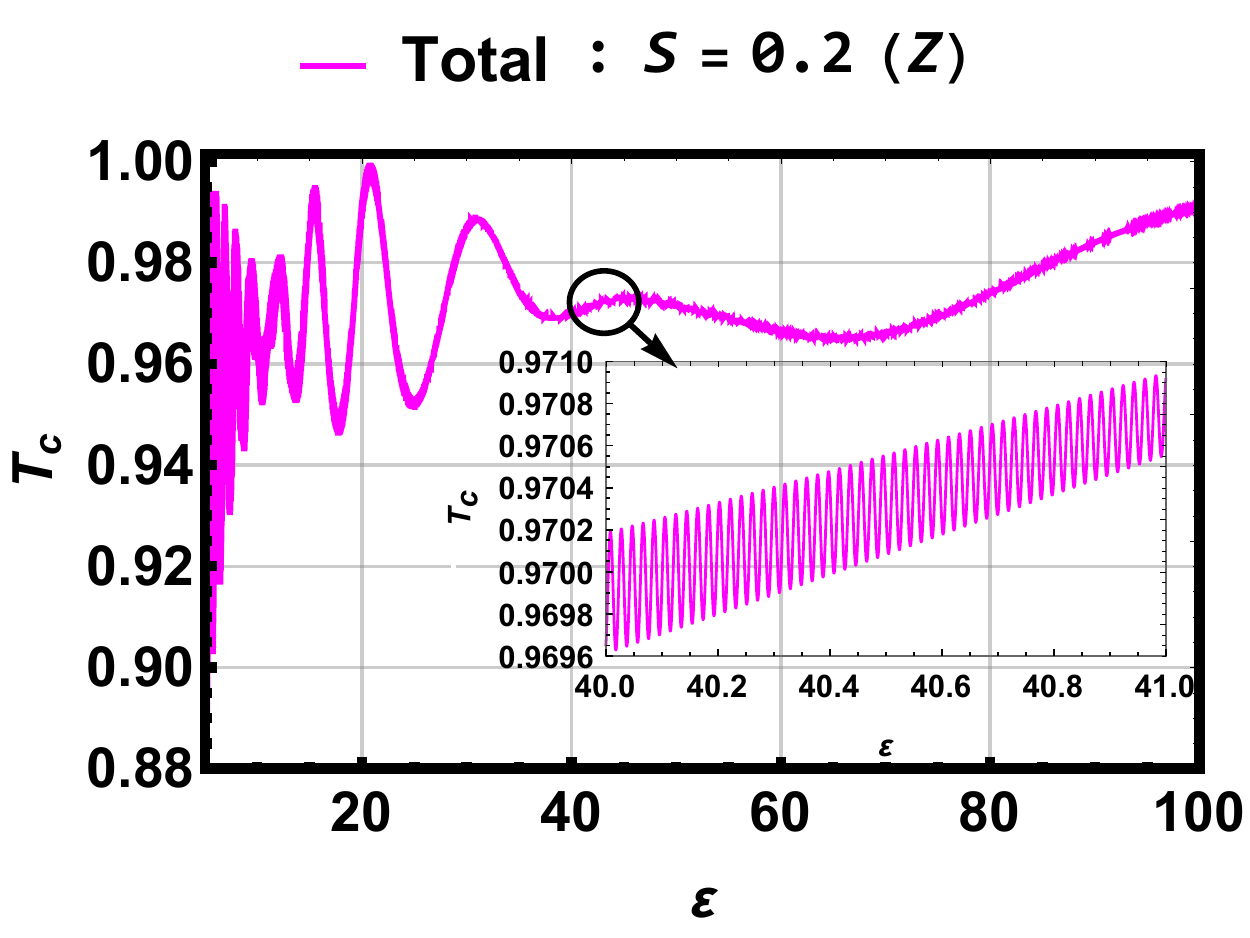}}}
  	\caption{(color online) Transmission probabilities  for central band $T_{0}$, first sidebands {\rr $T_{\pm1}$} and  $T_{c}$ versus the incident energy $\varepsilon$ for {$L/l_{B}=60$, $\tilde{F}=0.658$, $\tilde{\omega}=1$, $k_{y}l_B=1.5$ and $S=0.0$, $S=0.2$ (A), $S=0.2$ (Z).}}\label{fiB4}
  \end{figure}

To demonstrate the effect of strain magnitude on the transmission probabilities $T_c $, we plot in Fig. \ref{fiB4} the numerical results of $T_ 0 $, $T_{\pm1}$, and $T_{c}$ versus the incident energy $\varepsilon$ for $L/l_{B}=60$, $\tilde{F}=0.658$, $\tilde{\omega}=1$, $k_{y}l_B=1.5$.  Indeed, we observe that the transmission probabilities show an oscillatory manifestation for all channels, especially at lower incident energy. Further, {one sees that in Fig. \ref{fiB4}a  for $S=0$ the transmission through the central band is higher than that through the sidebands, which is consistent with the findings obtained in \cite{Xu5,Xu51}.}
On the other hand, we can clearly see that when the strain is applied along the armchair direction in Fig. \ref{fiB4}c,  the number of oscillations decreases dramatically and $T_ c $ becomes nearly linear by increasing the values of $ \varepsilon $. 
As shown in 
Fig. \ref{fiB4}e, the strain along the zigzag direction has a significant effect on all modes of $T_c$.
It not only sharply alters the amplitude of transmission, but it also changes the period of its oscillations. 
Furthermore, Figs. \ref{fiB4}(b,d,f)
show that the total transmission includes everything discovered for $T_ 0 $ and $T_{\pm1}$. When compared to the strainless case, $T_c$ displaces to the right in an armchair situation but to the left in a zigzag situation. Also, its oscillations are growing more rapidly, and this seems clear through the zoom that we have done in some zones. Therefore, we conclude that the strain exerted in different directions represents the opposite behavior in transmission probability.

\begin{figure}[ht]\centering
  	{	\subfloat[]{\includegraphics[width=0.45\linewidth, height=0.19\textheight]{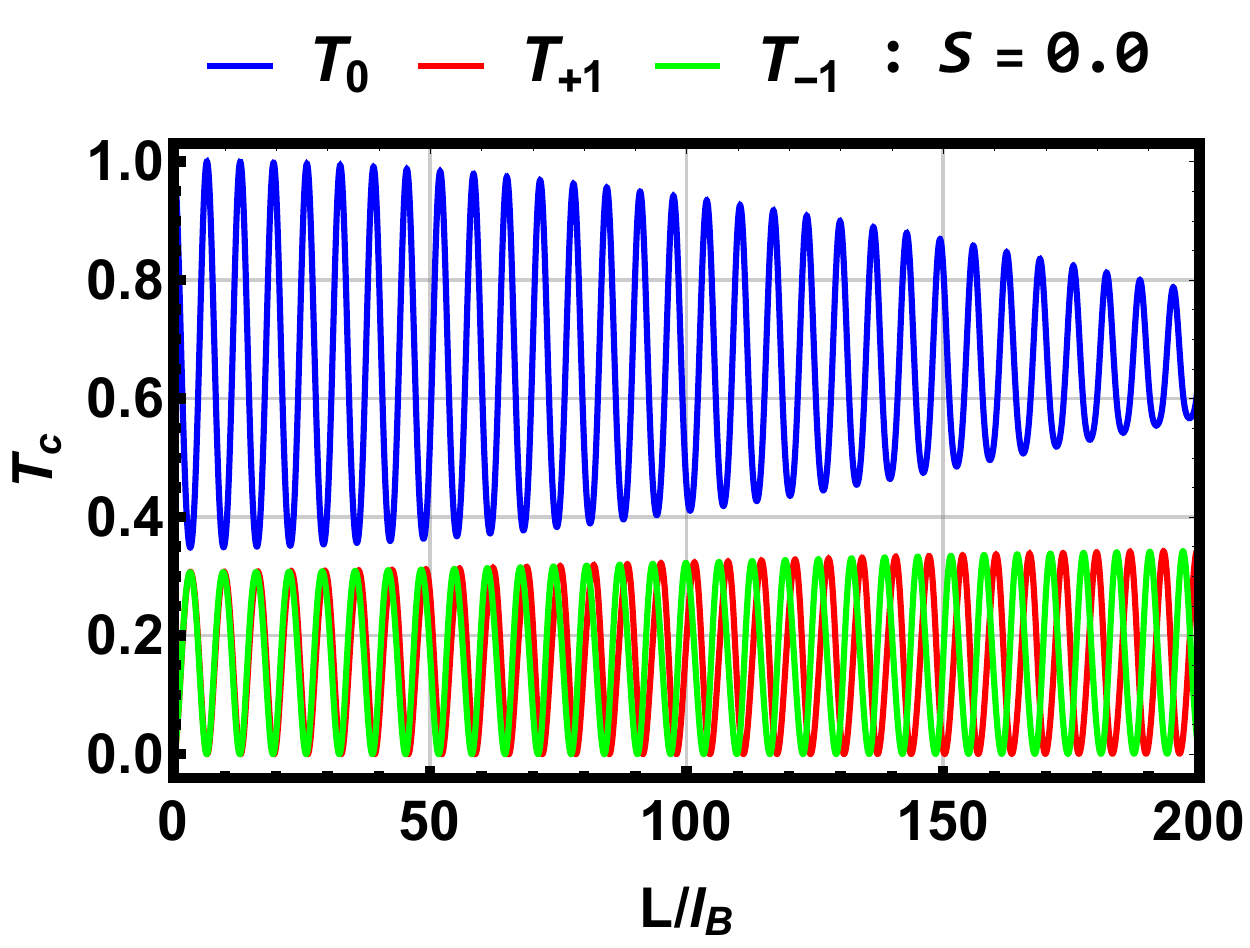}}}
  	{\subfloat[]{
  			\includegraphics[width=0.46\linewidth, height=0.19\textheight]{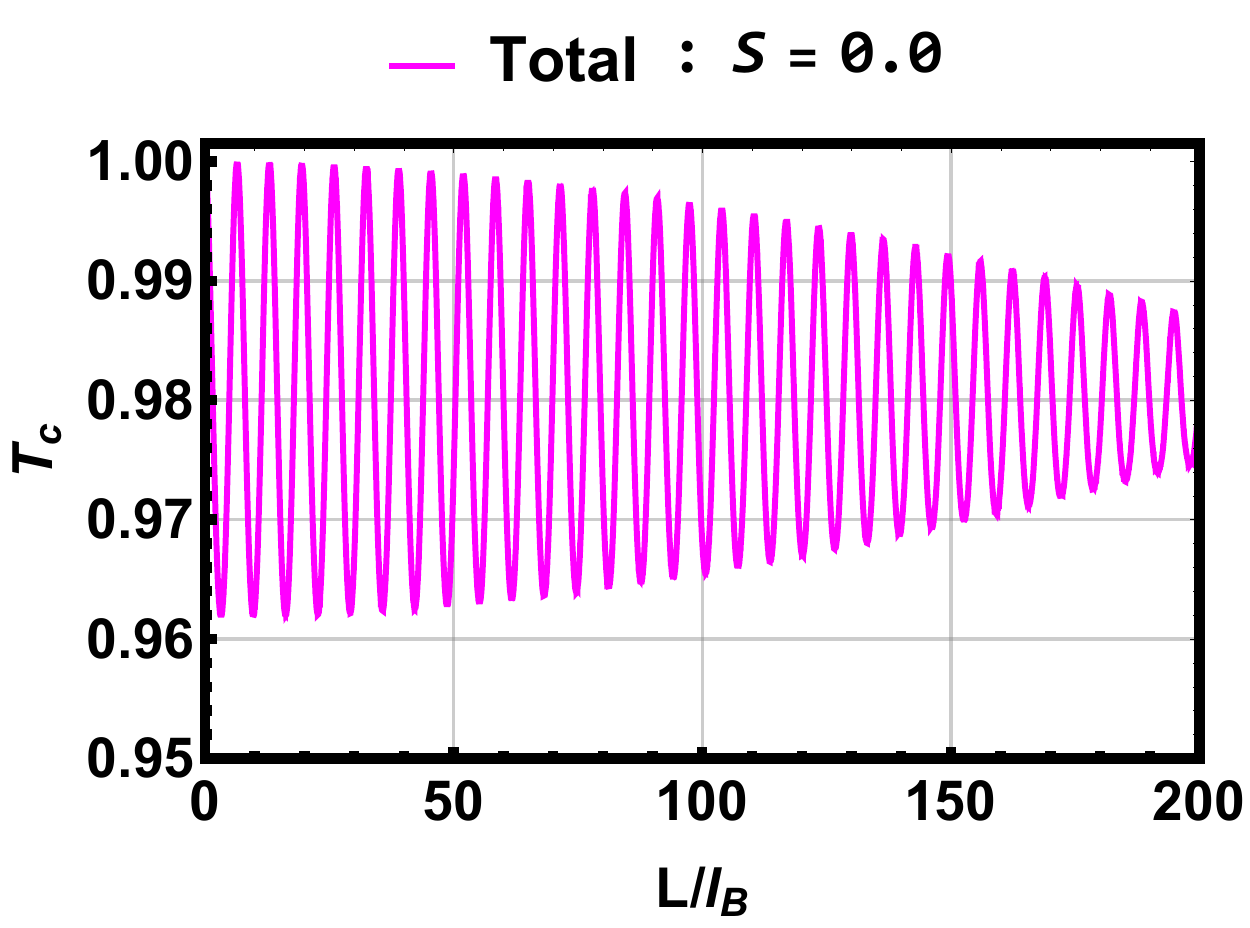}}}\\
  	{\subfloat[]{
  			\includegraphics[width=0.45\linewidth, height=0.19\textheight]{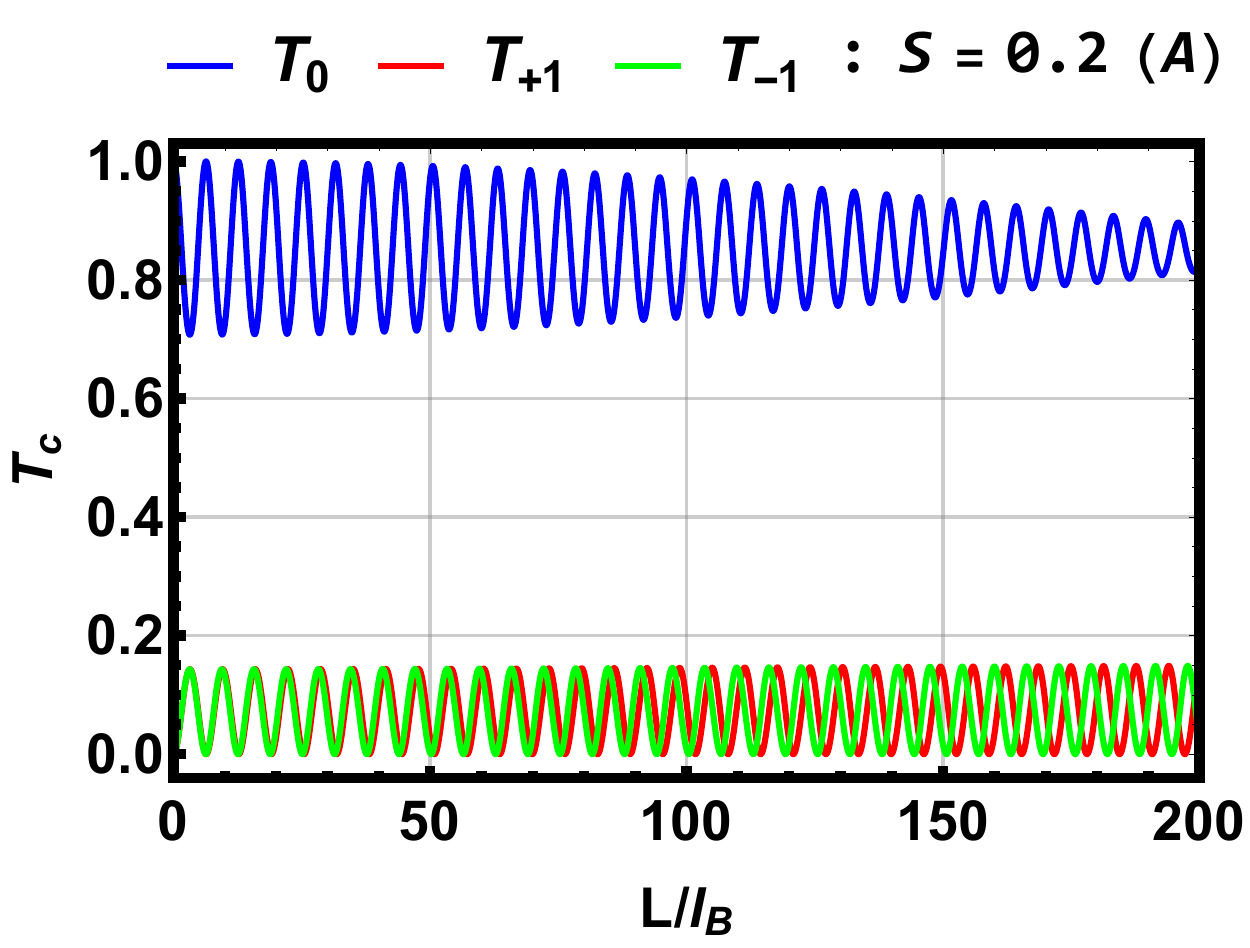}}}
  	{\subfloat[]{
  			\includegraphics[width=0.46\linewidth, height=0.19\textheight]{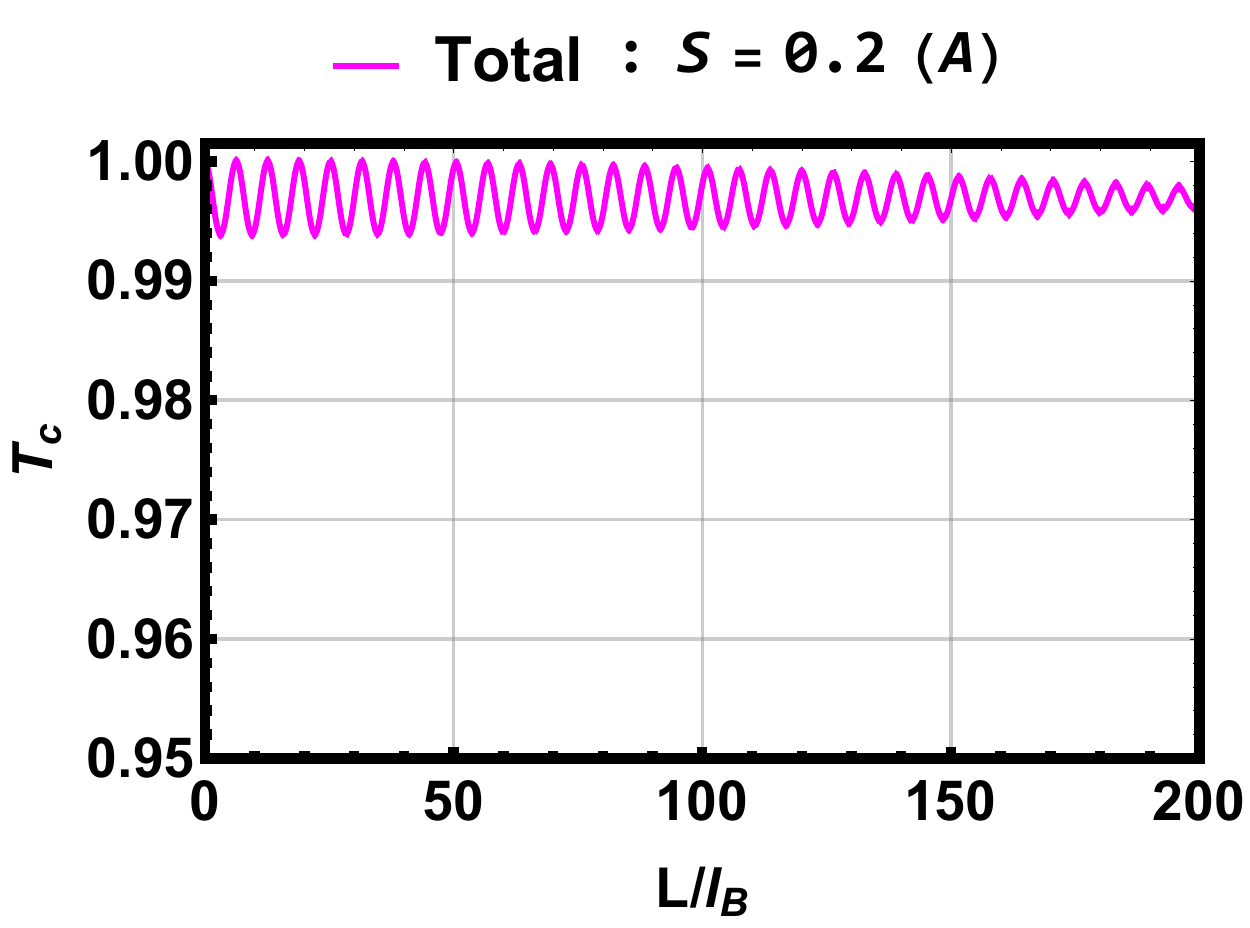}}}\\
  	{\subfloat[]{
  			\includegraphics[width=0.45\linewidth, height=0.19\textheight]{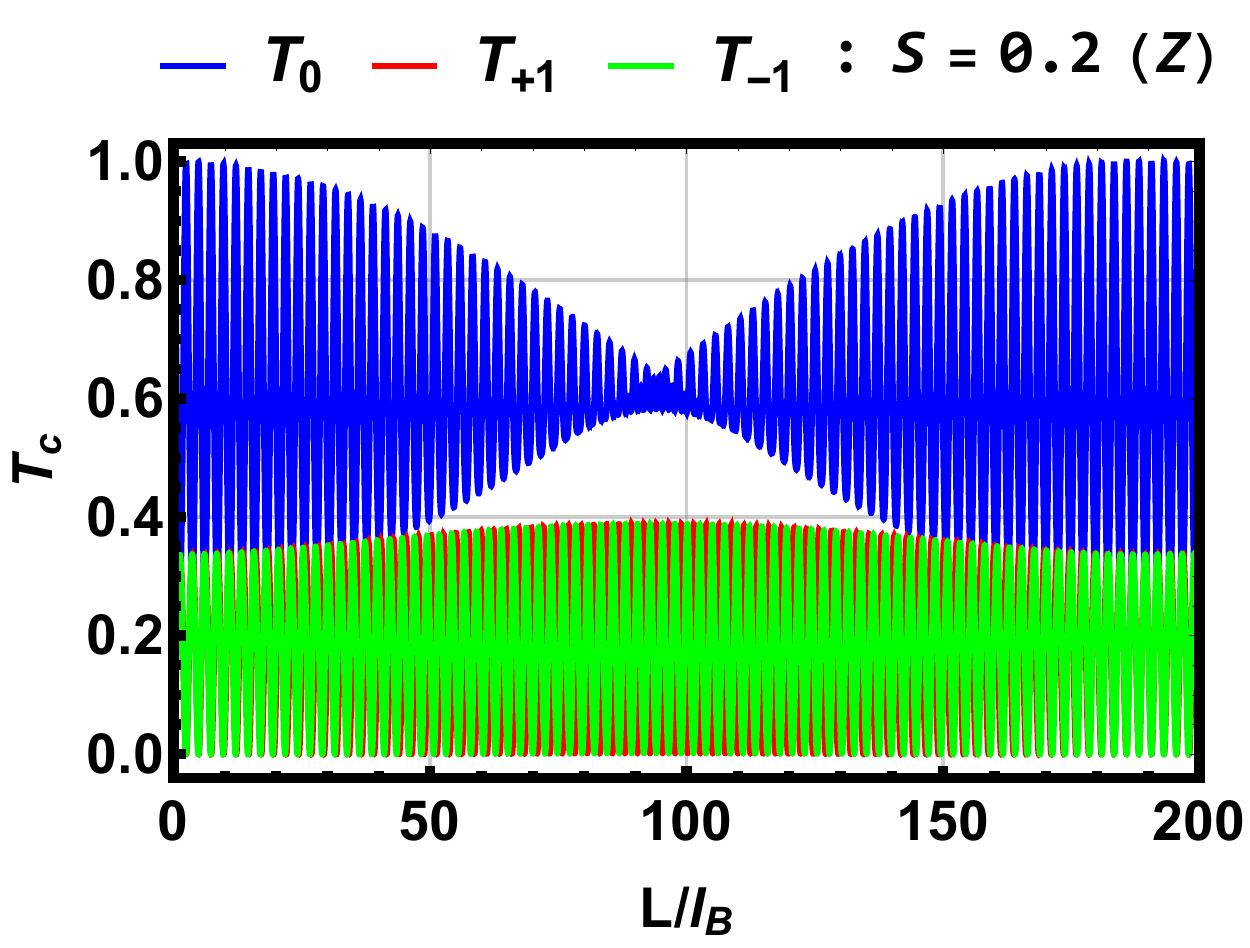}}}
  	{\subfloat[]{
  			\includegraphics[width=0.46\linewidth, height=0.19\textheight]{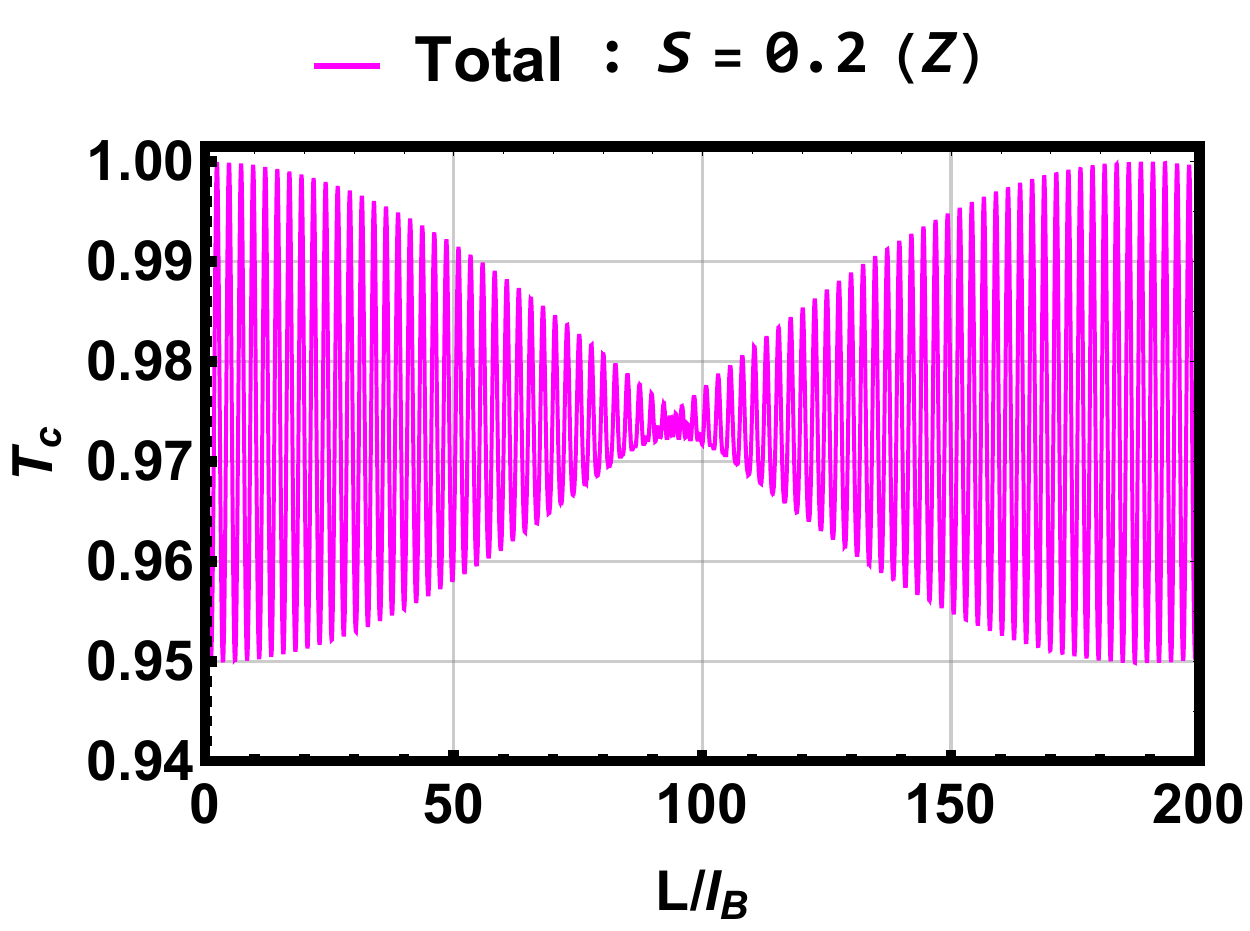}}}
  	\caption{(color online) Transmission probabilities  for central band $T_{0}$, first sidebands {\rr $T_{\pm1}$} and  $T_{c}$   versus the barrier width $L/l_{B}$ for {$\varepsilon=75$, $\tilde{F}=0.658$, $\tilde{\omega}=1$, $k_{y}l_B=1.5$ and $S=0.0$, $S=0.2$ (A), $S=0.2$ (Z).}}\label{fi41}
  \end{figure}

We display in Fig. \ref{fi41} the transmission probabilities for the central band $T_{0}$, first sidebands $T_{\pm1}$ and $T_{c}$  versus the barrier width $L/l_B$ with $\varepsilon=75$, $\tilde F=0.658$, $\tilde\omega=1. k_yl_B=1.5$, and  three strain values, $S=0$, armchair ($S=0.2$), and zigzag ($S=0.2$). 
{The first result observed in Fig. \ref{fi41}a is that the central band transmission begins at unity and swings periodically, but it gradually decreases for larger values of the barrier width, in contrast to the situation of an oscillating barrier in which the maxima of peaks remain constant in one, as reported in \cite{A7}.  In fact, the presence of the laser light in the intermediate region is the reason for this behavior.}
 Meanwhile, transmission for the other nearest sidebands begins at zero and continues until the values $L/l_B\lesssim30$ are reached.
{After that, it is no longer identical and takes the form of a sinusoidal function that increases for higher values of $L/l_B$.} When the strain is exerted along the armchair direction with $S=0.2$ as shown in Fig. \ref{fi41}c, we remark that the amplitude of transmission drops dramatically and its oscillations move to the up for $T_{0}$ and to the down for $T_{\pm1}$. 
 Conversely, for zigzag strain direction, we observe the appearance of more and more peaks in transmission, see 
 Fig.~\ref{fi41}e. Indeed, we observe symmetry in all transmission probability channels for $L/l_B\simeq95$. 
 {As shown in \cite{dx1}, the central band $T_0$ exhibits the characteristics of beating oscillations with oscillations of various frequencies.} 
 $T_{\pm 1}$ behaves similarly to $T_0$ 
  in Figs. \ref{fi41}(b,d,f), 
  except that it is translated and the number of oscillations increases significantly for strainless and zigzag cases but decreases significantly for armchair strain. Then, we stress that $T_c$ is highly dependent on the direction of applied strain and the barrier width.
  
\begin{figure}[ht]\centering
	{\subfloat[]{\includegraphics[width=0.45\linewidth, height=0.19\textheight]{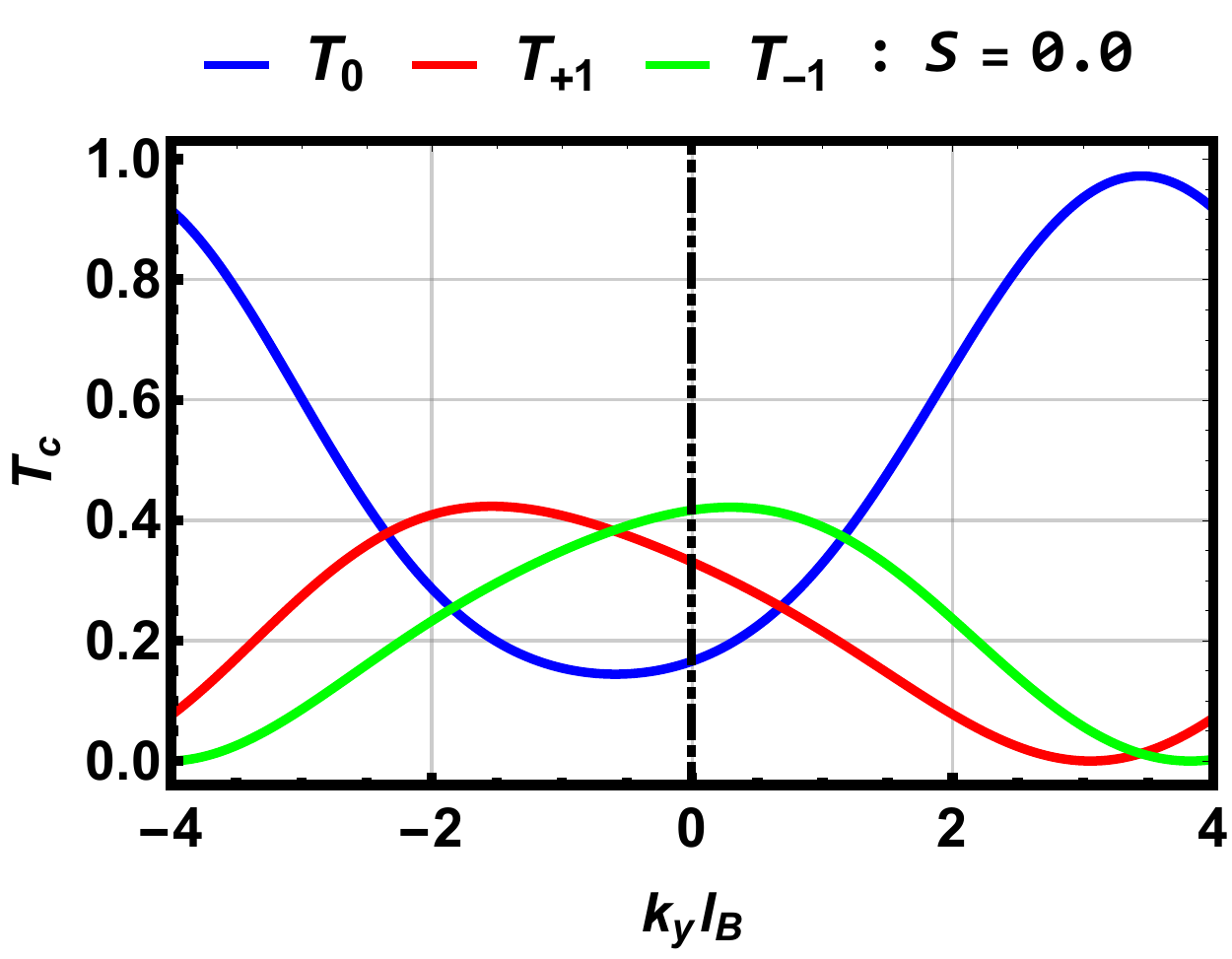}}}
	{\subfloat[]{
			\includegraphics[width=0.46\linewidth, height=0.192\textheight]{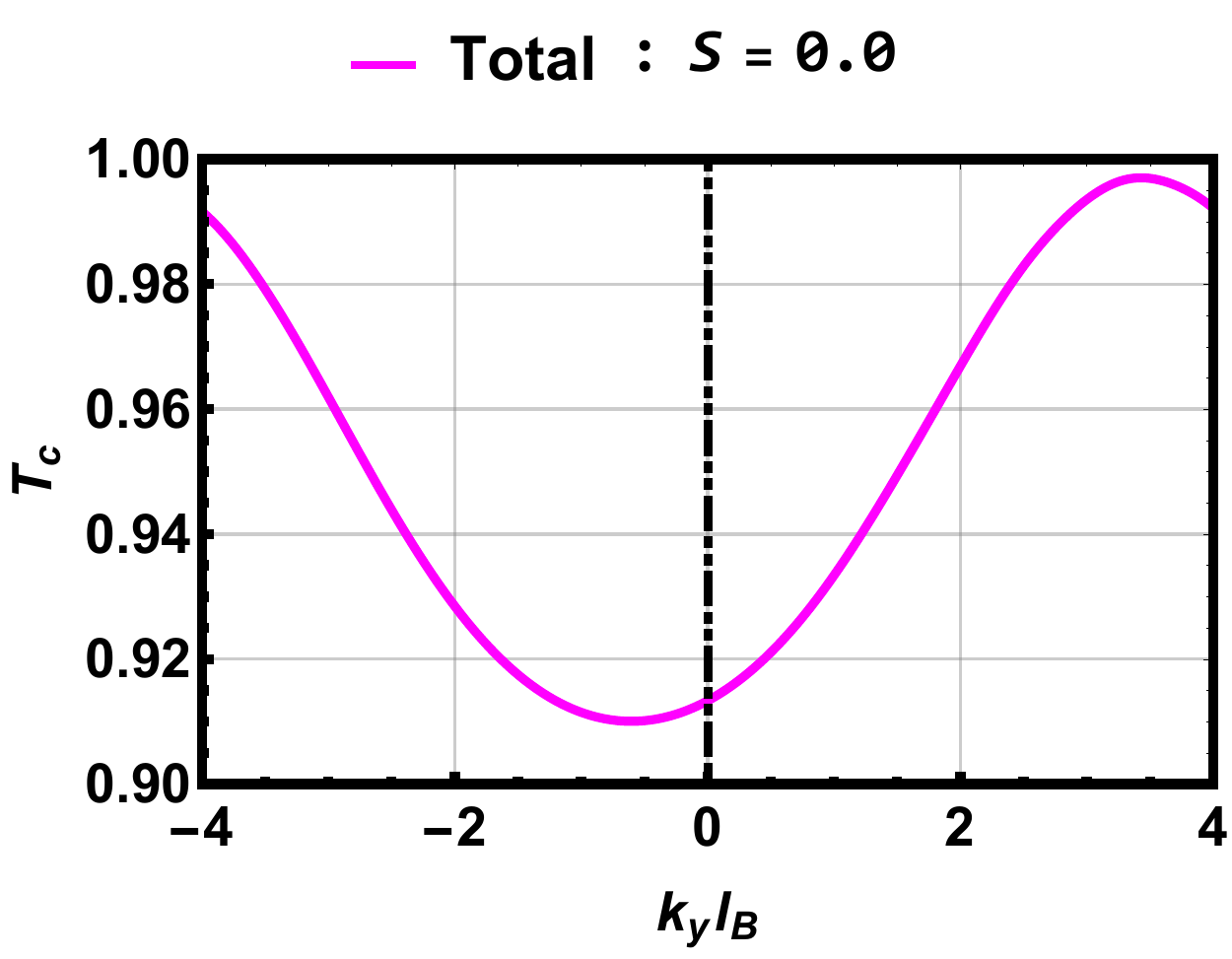}}}\\
	{\subfloat[]{
			\includegraphics[width=0.45\linewidth, height=0.19\textheight]{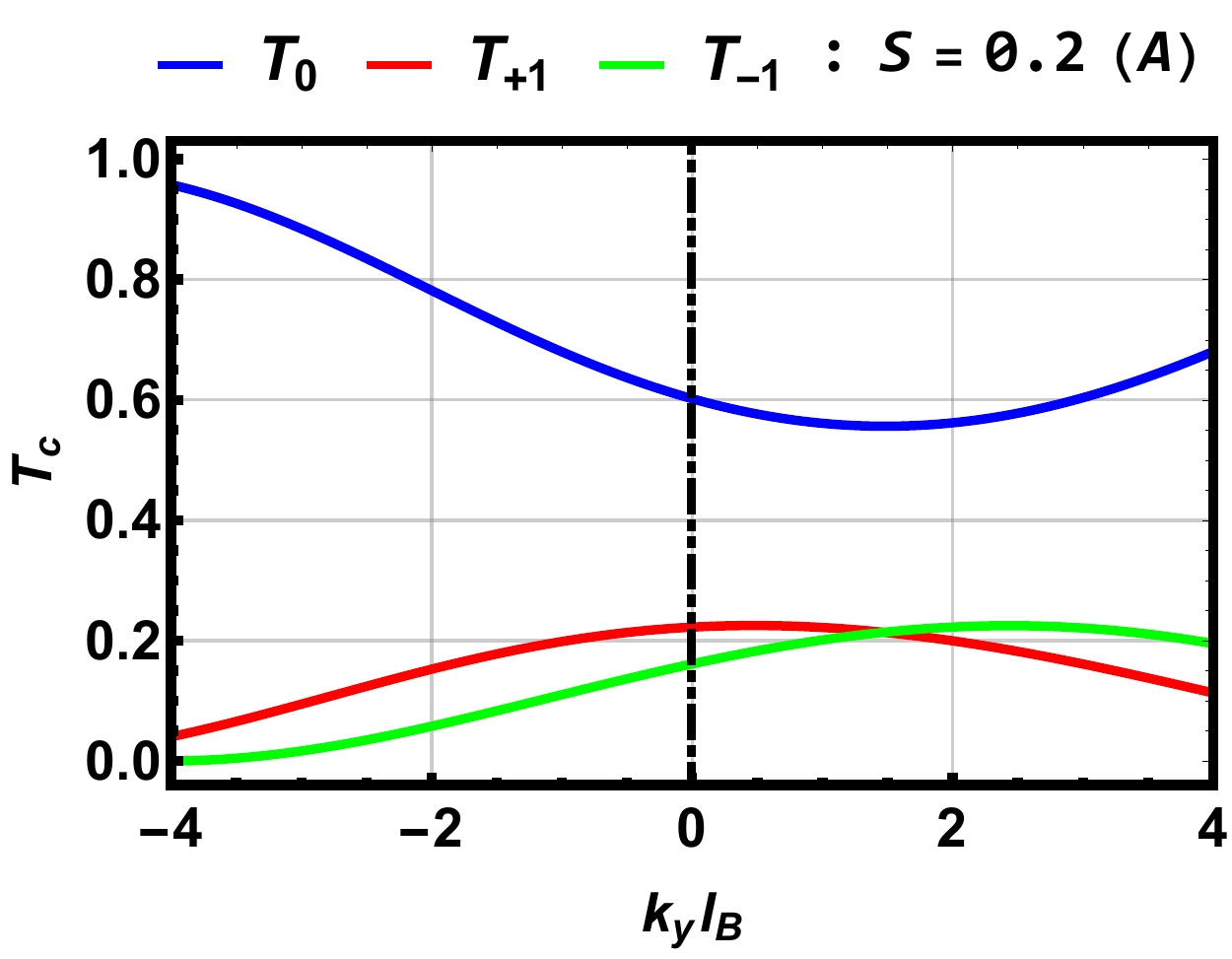}}}
	{\subfloat[]{
			\includegraphics[width=0.46\linewidth, height=0.192\textheight]{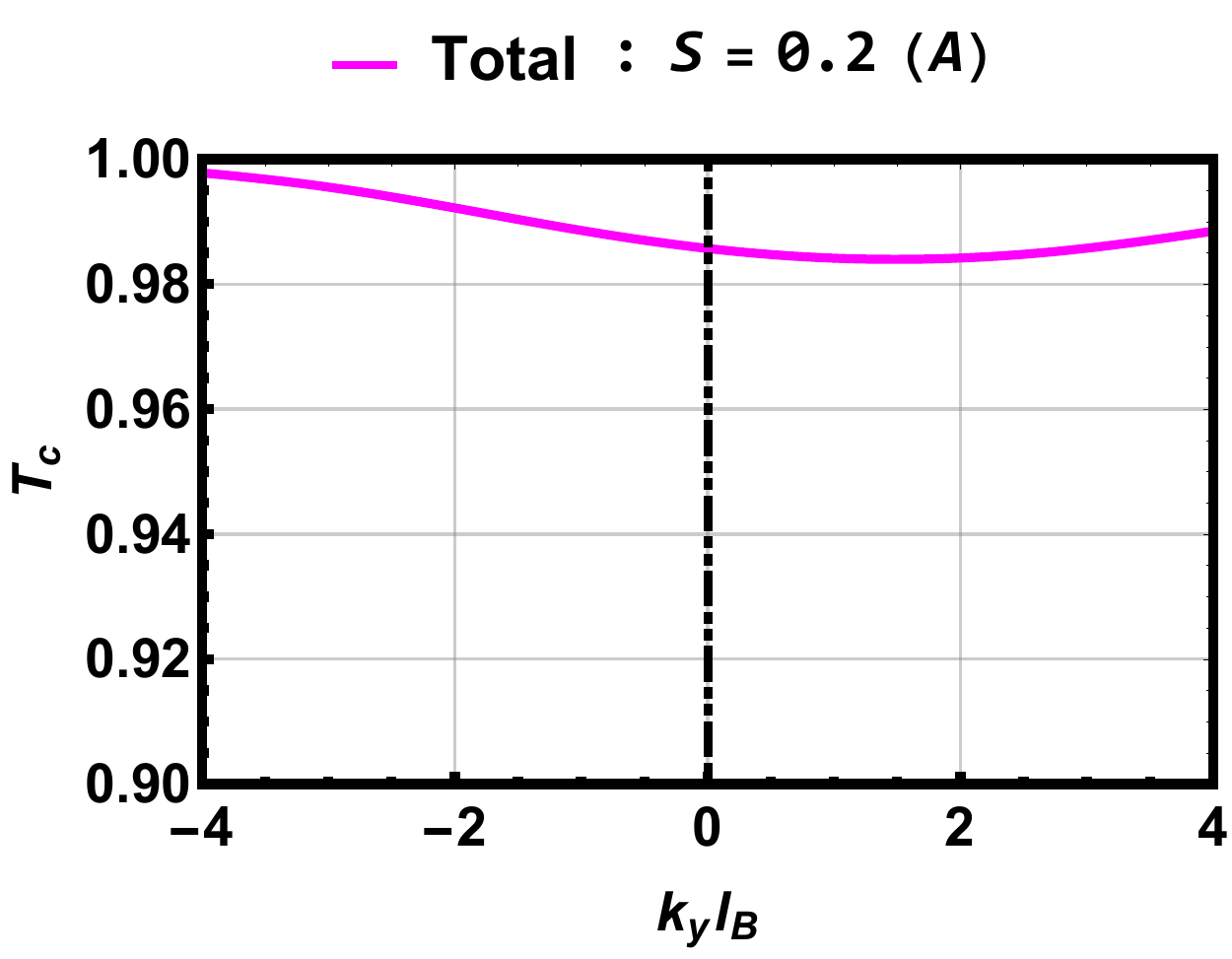}}}\\
	{\subfloat[]{
			\includegraphics[width=0.45\linewidth, height=0.19\textheight]{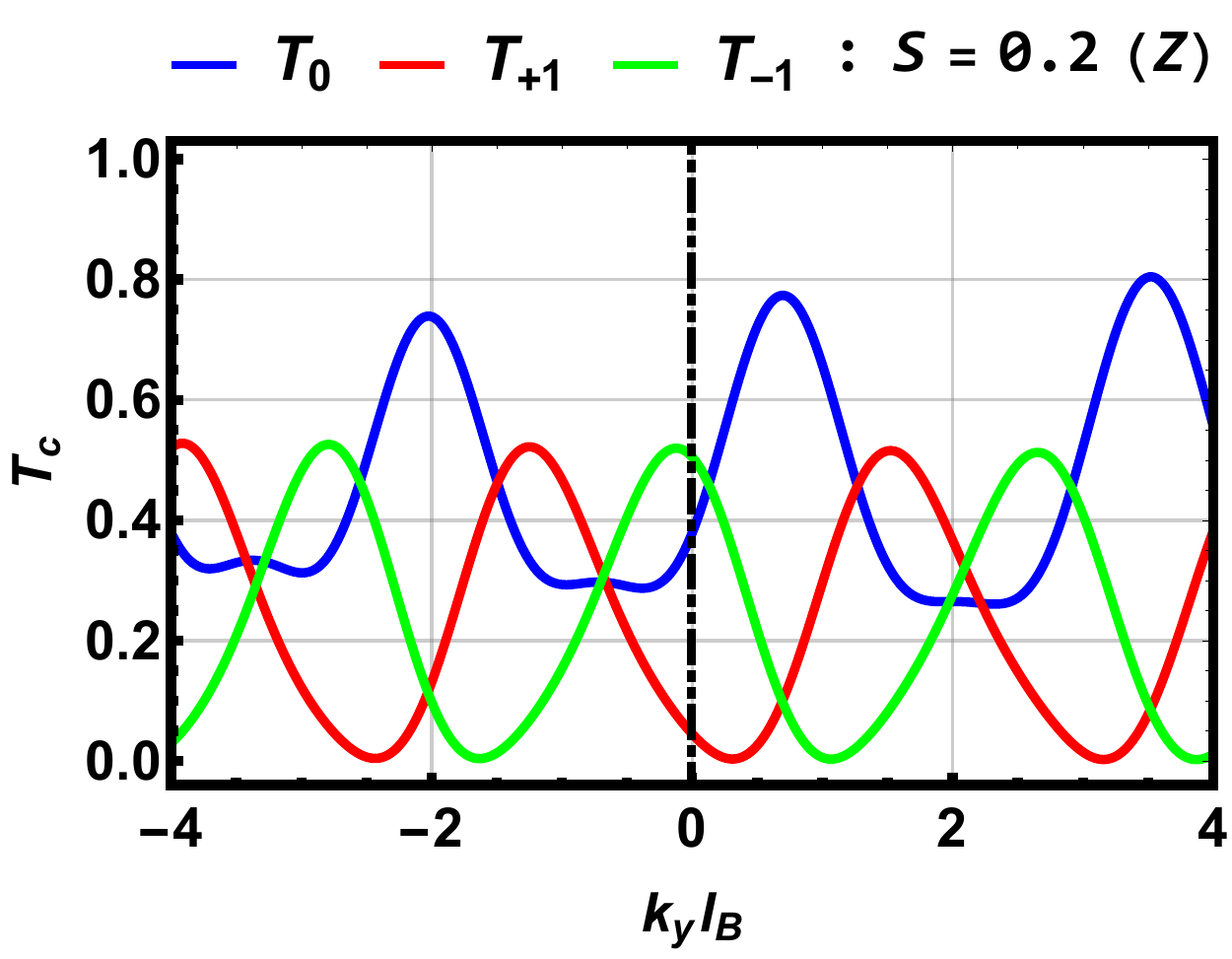}}}
	{\subfloat[]{
			\includegraphics[width=0.46\linewidth, height=0.192\textheight]{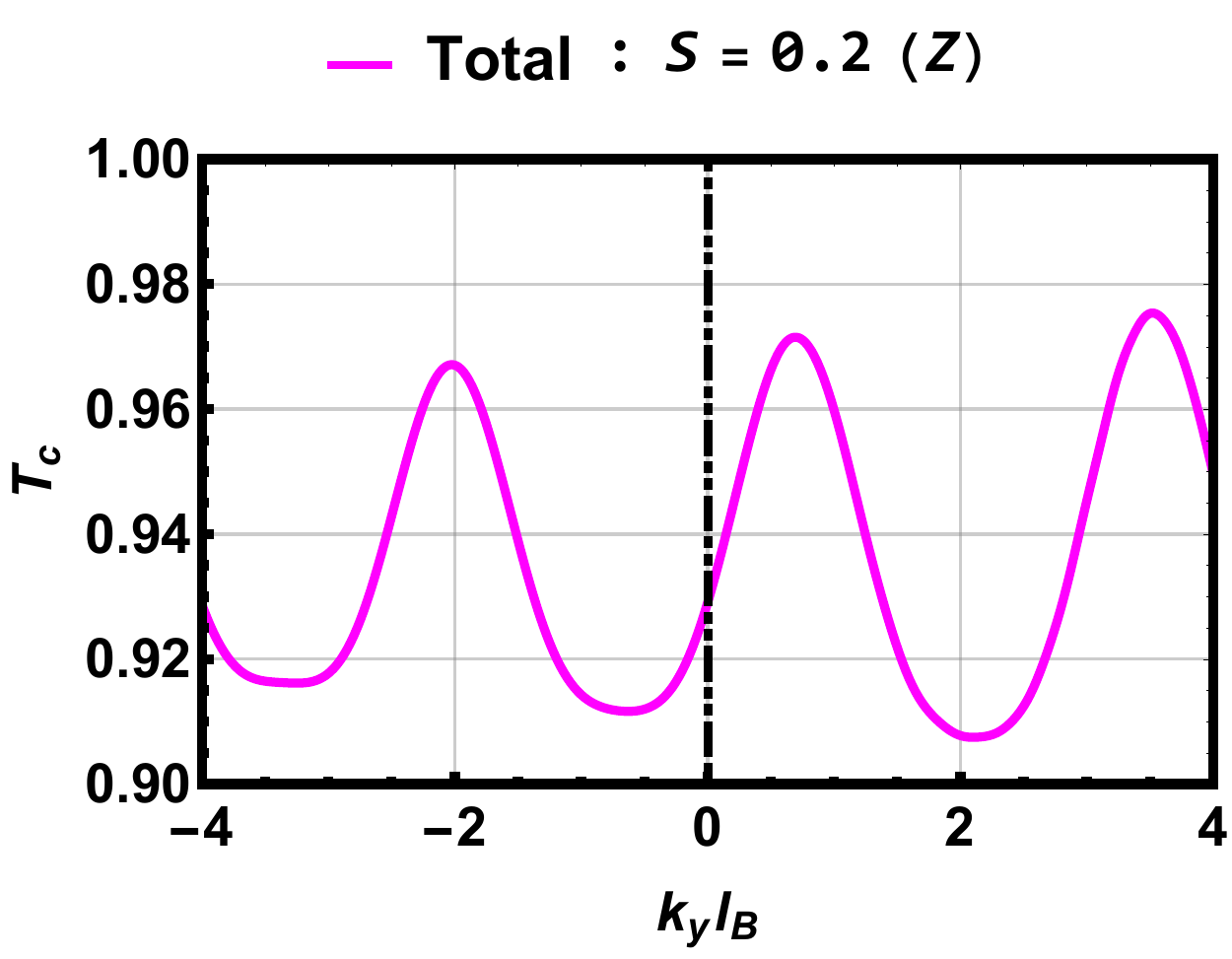}}}
	\caption{(color online) Transmission probabilities for central band $T_{0}$, first sidebands {$T_{\pm1}$} and  $T_{c}$   versus the wave vector component $k_{y}{l_B}$ for {$\varepsilon=75$,  ${L}/{l_B}=60$, $\tilde{F}=0.858$, $\tilde{\omega}=1$ and $S=0.0$, $S=0.2$ (A), $S=0.2$ (Z).}}\label{fi4}
\end{figure}

Fig. \ref{fi4} depicts 
the transmission probabilities  versus the wave vector component $k_ yl_ B$, with
$\varepsilon=75$, $L/l_B=60$, $\tilde F=0.858$, $\tilde\omega=1$, and  $S=0.0$, $S=0.2$ (A), $S=0.2$ (Z). 
According to Fig. \ref{fi4}a, $T_ l$ across the central band $(l=0)$ and the sidebands $(l=\pm 1)$ exhibits the oscillatory pattern between positive and negative values taken by $k_ yl_B$. 
{In contrast to the electrostatic barrier \cite{A1,A6}, the transmission is clearly asymmetric with respect to the $k_ yl_B$ sign. Also, as seen in \cite{str1}, the first sidebands at normal incidence, i.e., $k_yl_B=0$, are not identical.}
When the strain is along the armchair direction, the oscillations of $T_c$ disappear and its width dramatically broadens (see Fig. \ref{fi4}c). 
 Otherwise, there is a noticeable difference if the strain is along a zigzag direction, where the number of peaks grows increasingly for $T_{0}$ but sinusoidally for  $T_{\pm1}$, which is not the situation of armchair one. On the other hand, one can see in Figs. \ref{fi4}(b,d,f), 
 the total transmission behaves asymmetrically in all cases of deformation. 
 Additionally, it is noticed that $T_{c}$ is much less than unity at $k_{y}l_{B}=0$, and therefore the Klein tunneling phenomenon is suppressed when a high laser field is applied. Regarding the influence of strain in two different directions, we have the same characteristics as for previous figures. These findings reveal that the present system can be controlled by modifying the values of the wave vector component and distortion.
 
%%%%%%%%%%%%%%%%%%%%%%%%%%%%%%%%%%%%%%%%%%%%%%%%%%
\section{Conclusion}
%%%%%%%%%%%%%%%%%%%%%%%%%%%%%%%%%%%%%%%%%%%%%%%%%%%%%%
We have theoretically investigated the strain influence along armchair and zigzag directions on the transmission probabilities across magnetic barriers of type delta-function in monolayer graphene subjected to linearly polarized laser light. By resolving the Dirac equation, we have analytically determined the eigenspinors that correspond to each region. These have been utilized with the transfer matrix formalism and the current densities to evaluate the transmission on the different sidebands of Floquet as a function of a set of physical parameters characterizing our system, such as the strain amplitude, magnetic field, incident energy, laser field amplitude, frequency, barrier width, and the wave vector components.
Subsequently, we have discussed our numerical results concerning the transmission probabilities for three cases: strainless, strain along armchair, and strain along zigzag. Indeed, we have found that for strainless cases, the transmission shows oscillatory behavior in all channels. By changing the values of the barrier width and the incident energy, we have observed that the transmission via the central band oscillates decreasingly with different amplitudes. The transmission via first sidebands increases from zero and takes the pattern of a sinusoidal function that grows for higher barrier width but becomes nearly linear as we augment the incident energy.

On the other hand, when the strain magnitude is applied along the armchair direction, we have noticed that the transmission displaces to the right and the number of its oscillations reduces rapidly. Contrariwise, the strain along a zigzag direction not only greatly changes the amplitude of peaks but also amends their period. Additionally, we have found that the transmission displays a symmetry with increasing values of distortion and barrier width. Another intriguing result in the current study is that the irradiation of the powerful laser field prevents the appearance of the Klein tunneling effect in all situations of strain. {In summary, we found that the addition of a strain amplitude can be used to modulate transmission. These results suggest that future applications in electronics, magnetics, and photonics may be facilitated by being able to modify the mechanical properties of graphene.}

\end{document}